\documentclass[fleqn,usenatbib]{mnras}

\usepackage{newtxtext, newtxmath}

\usepackage[T1]{fontenc}

\DeclareRobustCommand{\VAN}[3]{#2}
\let\VANthebibliography\thebibliography
\def\thebibliography{\DeclareRobustCommand{\VAN}[3]{##3}\VANthebibliography}

\usepackage{graphicx}	
\usepackage{amsmath}	
\usepackage{amssymb}	
\usepackage{multicol}

\usepackage{physics}
\usepackage{xcolor}
\usepackage{mathrsfs}
\usepackage{tabularx}
\usepackage[title]{appendix}
\usepackage[tight]{subfigure}
\allowdisplaybreaks
\usepackage{CJKutf8}

\newcommand{\sign}[1]{\mathrm{sgn}\qty(#1)}
\graphicspath{{./}{figures/}}

\title{Fluid Simulations of Cosmic Ray Modified Shocks}

\author[Tsung et al.]{
Tsun Hin Navin Tsung,$^{1}$\thanks{E-mail: ttsung@ucsb.edu}
S. Peng Oh,$^{1}$
Yan-Fei Jiang(姜燕飞)$^{2}$
\\
$^{1}$Dept. of Physics, University of California, Santa Barbara, CA 93106, USA\\
$^{2}$Center for Computational Astrophysics, Flatiron Institute, New York, NY 10010, USA\\
}

\date{Accepted XXX. Received YYY; in original form ZZZ}

\pubyear{2020}

\begin{document}
\begin{CJK*}{UTF8}{gbsn}
\label{firstpage}
\pagerange{\pageref{firstpage}--\pageref{lastpage}}
\maketitle

\begin{abstract}
We consider cosmic ray (CR) modified shocks with both streaming and diffusion in the two-fluid description. Previously, numerical codes were unable to incorporate streaming in this demanding regime, and have never been compared against analytic solutions. First, we find a new analytic solution highly discrepant in acceleration efficiency from the standard solution. It arises from bi-directional streaming of CRs away from the subshock, similar to a Zeldovich spike in radiative shocks. Since fewer CRs diffuse back upstream, this results in a much lower acceleration efficiency, typically $\lesssim 10\%$ as opposed to $\gtrsim 50\%$ found in previous analytic work. At Mach number $\gtrsim 10$, the new solution bifurcates into 3 branches, with efficient, intermediate and inefficient CR acceleration. Our two-moment code \citep{jiang18} accurately recovers these solutions across the entire parameter space probed, with no ad hoc closure relations. For generic initial conditions, the inefficient branch is the most robust and preferred solution. The intermediate branch is unstable, while the efficient branch appears only when the inefficient branch is not allowed (for CR dominated or high plasma $\beta$ shocks). CR modified shocks have very long equilibration times ($\sim 1000$ diffusion time) required to develop the precursor, which must be resolved by $\gtrsim 10$ cells for convergence. Non-equilibrium effects, poor resolution and obliquity of the magnetic field all reduce CR acceleration efficiency. Shocks in galaxy scale simulations will generally contribute little to CR acceleration without a subgrid prescription. 
\end{abstract}

\begin{keywords}
Cosmic Rays -- Shock Waves -- MHD
\end{keywords}



\section{Introduction} \label{sec:introduction}

Cosmic rays (CR) are close to energy equipartition with thermal gas in the local ISM, and have been observed in many astrophysical scenarios. They are now thought to be dynamically important to galaxy evolution, both in providing non-thermal support to the CGM gas and in driving a wind that initiates a feedback cycle (e.g., see \citealt{zweibel17} for a recent review), which has become the focus of intense study by numerous groups in recent years. It has even been suggested that the circumgalactic medium is CR dominated \citep{ji19}. CRs are believed to be accelerated at shocks to high energies through DSA (Diffusive Shock Acceleration). Test particle theories developed in the 1970s \citep{krymsky77, axford77, bell78a, blandford78} were instrumental in explaining the observed power law in CR energy. It was later realized that CR coupling to the background thermal gas through plasma instabilities can affect the acceleration efficiency by generating a shock precursor where upstream thermal particles can be decelerated, compressed and scattered, thus facilitating further acceleration \citep{drury81}. The two-fluid model and Monte-Carlo simulation were two common methods utilized to study this nonlinear behavior. These variant models all point to the same conclusion, that the non-linear modification of the shock by CRs is substantial. 

Magnetic field amplification due to compression, baroclinic vorticity and plasma instabilities can be dynamically important too, and has been seen in X-ray observations \citep{ballet06, morlino10}.  With the growth of computational power it became possible to perform PIC/hybrid simulations which capture the most important microphysics of CR shock acceleration, including various kinetic instabilities and their non-linear evolution into turbulence (e.g., \citealt{caprioli14a}). These simulations continue to show that shock acceleration is very efficient. 

In this paper, we study CR modified shocks in the two fluid approximation ubiquitously used in galaxy formation simulations of CR feedback. CRs couple with the background gas through the streaming instability \citep{kulsrud69}. In this instability, CR bulk drifting at velocity greater than the local Alfven wave speed ($v_D > v_A$) excites magnetic waves which gyro-resonantly scatter the CR, effectively locking the drift motion of the CR to the local wave frame ($\qty(v_D - v_A)/v_A \ll 1$), causing it to `stream' along the magnetic field at the Alfven speed down the CR pressure gradient, i.e.,

\begin{equation}
    \vb{v}_s = -\vb{v}_A\frac{\vb{B}\cdot\nabla P_c}{\abs{\vb{B}\cdot\nabla P_c}}. \label{eqn:streaming_speed}
\end{equation}

This collective streaming causes energy transfer from CR to the gas at the volumetric rate of $\vb{v}_s\cdot\nabla P_c$. In steady state, wave growth is balanced by various damping mechanisms (e.g., see \citealt{wiener13}). The finite scattering rate of CRs means that they are not perfectly locked to the Alfven frame; slippage with respect to the Alfven frame is expressed in terms of an anisotropic diffusive flux $\bar{\kappa} \nabla P_c$, 
where $\bar{\kappa}$ is dependent on the CR energy spectrum, the various plasma parameters and the damping mechanisms at play. We forgo these complications and assume the diffusion coefficient is constant in time and space though our work can be extended to account for a more detailed treatment of diffusion.

The two fluid treatment was historically the first method used to study CR modified shocks. However, it has several shortcomings. Since momentum information is integrated out, CR pressure and energy (which are moments of the full distribution function) have to be related by an equation of state, with adiabatic index $\gamma_{\rm c} = 1+ P_{\rm c}/E_{\rm c}$ which is usually assumed to be constant, $\gamma_{\rm c} =4/3$. In reality, $\gamma_{\rm c}$ depends on the detailed shape of the distribution function and evolves continuously from $5/3$ to $4/3$ as particles are accelerated. Shock structure, compressibility and acceleration efficiency are all sensitive to assumptions about the adiabatic index \citep{achterberg84, duffy94}. Similarly, the diffusion coefficient $\bar{\kappa}$ is averaged over the CR spectrum. Furthermore, it is not self-consistently calculated\footnote{The calculation of the diffusion coefficient itself requires calculating wave growth by the resonant streaming instability \citep{kulsrud69}, the current-driven non-resonant Bell instability \citep{bell04}, as well as associated damping mechanisms. Our current study essentially assumes that waves are strongly damped, although kinetic simulations show that waves can be amplified to the non-linear regime \citep{caprioli14b}, which facilitates CR scattering.}. In general it should evolve with the time-dependent distribution function. In particular, since generically $\kappa (p)$ rises with energy, this can lead to a CR flux dominated by the highest-energy particles; in this case a steady state shock structure no longer exists. In this paper, we simply assume a constant, time-steady $\bar{\kappa}$ (and hereafter drop the overbar). Finally, the standard CR hydrodynamic equations ignore microscopic physics such as thermal injection and MHD wave growth which PIC and hybrid simulations take into account\footnote{They can potentially be modified to include such physics; we implement a very simplified prescription for thermal injection (\S\ref{sec:thermal}), and one can also analytically model wave growth \citep{caprioli08, caprioli09}}. 

Given these serious shortcomings, it may seem a step backwards to simulate CR modified shocks using the two-fluid approach. Certainly, if our main interest is understanding CR acceleration at shocks, then PIC and hybrid simulations are unquestionably the tools of choice. However, there are still compelling reasons for two-fluid CR shock simulations: 

{\it Code testing.} In recent years, as interest in the role of CRs in galaxy formation has rapidly grown, many new codes for simulating CR transport in the two fluid approximation have been developed (CR streaming with regularization \citep{sharma09}; ENZO \citep{salem14}; AREPO \citep{pfrommer17}; GIZMO \citep{chan19}; GADGET-2 \citep{pfrommer06}; RAMSES \citep{booth13}; FLASH \citep{yang12} among others). These must be subjected to a battery of tests to ensure they are correctly solving the CR transport equations. Perhaps the most demanding test for such codes are CR shocks; this is also one of the few regimes where analytic solutions exists. However, to date codes have only been compared against analytic solutions in the purely advective regime, with both CR streaming and diffusion turned off. Even in this restricted regime, numerical methods do not appear to be robust. When the postshock CR pressure is a small fraction of the gas pressure, simulations appear to agree with existing analytic solutions \citep{pfrommer06, pfrommer17, dubois19}. However, once this is no longer true, outcomes are non-unique and dependent on numerical method such as discretization, time-steppping, spatial reconstruction, and CFL number \citep{kudoh16, gupta19}. This was attributed to the fact that the equations can no longer be written in conservative form, due to the presence of a source term involving spatial derivatives. It was therefore suggested that additional assumptions are required at CR shocks to achieve closure, such as constant CR entropy across the shock \citep{kudoh16}, or a priori prescription of the post-shock CR pressure \citep{gupta19}. We shall clarify this situation by showing that such potentially unphysical assumptions are unnecessary in the full problem where CR transport (diffusion and streaming) is considered. 

Even more pressing is the need to compare codes with CR streaming to analytic solutions. In the past, simulations with CR streaming have been afflicted by severe grid-scale instabilities due to the requirement that CRs can only stream down their gradient \citep{sharma09}. The only known cure, adding artificial diffusion, led to severe time-step requirements ($\Delta t \propto (\Delta x)^2$ as well as dependence on the adopted smoothing parameter). Thus, simulations with CR streaming (and particularly CR shocks with streaming) were infeasible. These problems were resolved with a new two moment method for CR transport \citep{jiang18}, which has no arbitrary smoothing and only linear time-step scaling with resolution ($\Delta t \propto \Delta x$); since then similar formulations (albeit with some important differences) have been proposed \citep{chan19, thomas19} and employed in galaxy formation simulations. For instance, \citet{thomas19} claim that expansion to $\mathcal{O}(v_{\rm A}^2/c^2)$ is necessary, but did not present a specific scenario demonstrating this claim. No codes to date have been compared against existing analytic solutions with streaming \citep{volk84}. We shall show that these old analytic solutions are in fact incomplete, and develop a new set of solutions. The \citet{jiang18} method matches the new analytic solutions we develop. 

{\it CR shocks in galaxy formation simulations.} Another compelling motivation to understand CR shocks in the two fluid approximation is that at present it is the only one used in galaxy formation simulations; no other method has been shown to be feasible. Shocks are also omni-present in such simulations, and it is important to understand the mutual interaction and impact of CRs on shocks and vice-versa, particularly in the presence of CR streaming. For instance, it is usually prescribed in cosmological simulations that $f_{\rm CR} \sim 10\%$ of supernova energy is injected into CR (via a subgrid recipe) and that most of the CRs in the simulation comes from this source. However, in a two fluid code, shocks will enhance CR energy density. Thus, shocks generated by e.g. SNe blast waves, galactic wind termination shocks (e.g., \citealt{bustard17}), and structure formation shocks may produce CRs in excess of that from sub-grid injection recipes, and also alter the spatial distribution of CRs. It is important to understand this effect and its dependence on numerical resolution. The simulation results must also be checked to ensure they make physical sense (for instance, that CR acceleration efficiencies are not wildly discrepant with PIC simulations), given the approximations inherent in the two fluid method. It is also important to understand how CRs affect shock jump conditions (e.g., compression ratios, which is {\it increased} in the presence of CRs), and whether the simulations are handling this correctly. Only by doing so can we assess whether the astrophysical impact of shocks is correctly handled, and the robustness of observational predictions which depend on conditions at the shock (e.g., radio relics; \citealt{botteon20}).  

The outline of this paper is as follows. In \S\ref{sec:analytics}, we develop analytic solutions for CR modified shocks, and in particular a new solution which takes bi-directional streaming into account. In \S\ref{sec:simulation}, we show simulation results and compare them to the analytic solution. This is followed by a study on the equilibration time, resolution dependence, effect of oblique magnetic fields and injection. We conclude in \S\ref{sec:conclusion}.

\section{Analytics} \label{sec:analytics}

\subsection{Governing Equations} \label{subsec:assumptions}

Our analytic study follows the treatment by \citet{volk84}, hereafter VDM84, of CR shocks with streaming, but with some important modifications. We consider 1D adiabatic, non-relativistic, steady-state shocks in the two-fluid approximation. As noted in the Introduction, we do not assume any injection of CRs from the thermal pool; we simply assume a non-zero upstream CR pressure. With a shock finding algorithm, it is possible to include prescriptions for thermal injection (e.g., \citealt{pfrommer17}), but we eschew this for the sake of simplicity. At high Mach numbers, it has been suggested that the acceleration efficiency is independent of injection \citep{eichler79, ellison84, achterberg84, falle87, kang90}. We also ignore magnetic field amplification and subsequent back-reaction on the shock, which can alter compressibility and hence CR acceleration efficiency \citep{caprioli08, caprioli09}. This is standard in the two fluid formalism. 

The time-dependent equations two fluid equations we solve in our 1D numerical simulations are \citep{jiang18}: 
    
\begin{align}
    \pdv{\rho}{t}& + \div(\rho\vb{v}) = 0 \nonumber \\
    \pdv{\vb{v}}{t}& + \vb{v}\cdot\nabla\vb{v} = -\frac{1}{\rho}\nabla P_g + \frac{1}{\rho}\vb{\sigma}_c\cdot\qty[\vb{F}_c - \vb{v}\qty(E_c + P_c)] \nonumber \\
    \pdv{P_g}{t}& + \vb{v}\cdot\nabla P_g + \gamma_g P_g\div{\vb{v}} = \qty(\gamma_g - 1)\vb{v}_s\cdot\vb{\sigma}_c\cdot\qty[\vb{F}_c - \vb{v}\qty(E_c + P_c)] \nonumber \\
    \pdv{P_c}{t}& + \qty(\gamma_c - 1)\div{\vb{F}_c} = -\qty(\gamma_c - 1)\qty(\vb{v} + \vb{v}_s)\cdot\vb{\sigma}_c\cdot\qty[\vb{F}_c - \vb{v}\qty(E_c + P_c)]  \nonumber \\
    \frac{1}{c^2}\pdv{\vb{F}_c}{t}& + \nabla P_c = -\vb{\sigma}_c\cdot\qty[\vb{F}_c - \vb{v}\qty(E_c + P_c)] \label{eqn:two_moment}
\end{align}

where subscripts $g$ and $c$ denotes the gas and CR respectively; $F_c$ denotes the CR flux. The interaction coefficient tensor is:

\begin{equation} 
\sigma_{c}^{-1} = \sigma_{c}^{\prime \, -1} + \frac{\qty(E_c + P_c)}{\abs{\vb{B}\cdot\nabla P_c}} \vb{B}\vb{v}_A \label{eqn:sigma}
\end{equation}

where $\sigma_{c}^{\prime} = \qty(\gamma_c - 1)/\kappa$, and $\kappa$ is the customary CR diffusion coefficient. There are 5 time-dependent PDEs for the 5 variables $\rho,v,P_g,P_c,F_c$. Note the presence of source terms in the equations, indicating momentum and energy exchance between the gas and CRs. Total momentum and energy are conserved, since the source terms for gas and CRs are equal and opposite. However, the system of equations as a whole is not conservative, due to these source terms. Thus, conservation laws alone will not close the system. 

In our 1D formulation, the B-field is parallel to the shock propagation direction and magnetic pressure/tension is ignored. In steady state, conservation of mass, momentum and energy gives:

\begin{gather}
    \rho v = \text{const}, \label{eqn:mass_conv} \\
    \rho v^2 + P_g  + P_c = \text{const}, \label{eqn:mom_conv} \\
    \rho v \qty(\frac{1}{2}v^2 + \frac{\gamma_g}{\gamma_g - 1}\frac{P_g}{\rho}) + F_c = \text{const}, \label{eqn:energy_conv}
\end{gather}

where all quantities are measured in the shock frame. This is supplemented by the steady-state CR energy equation: 

\begin{equation}
    \dv{F_c}{x} = \qty(v + v_s)\dv{P_c}{x}, \label{eqn:cr_energy}
\end{equation}

where the steady state CR flux is: 

\begin{equation}
    F_c = \frac{\gamma_c}{\gamma_c - 1}\qty(v + v_s)P_c - \frac{\kappa}{\gamma_c - 1}\dv{P_c}{x} \label{eqn:cr_flux}
\end{equation}

Equation \ref{eqn:cr_energy} captures energy transferred from CRs to the gas, either by mechanical work done ($v \cdot \nabla P_{\rm c}$), or heating ($v_s \cdot \nabla P_c$). Transport by streaming and diffusion are captured respectively by the first and second terms on the RHS of eqn.\ref{eqn:cr_flux}. VDM84 assumed that CRs only stream towards the upstream. However, this assumption is unclear downstream given equation \ref{eqn:streaming_speed}; CRs can only stream down their gradient. We therefore restrain from presupposing a CR streaming direction. The direction will become clear as we go along. In the following, we take $\gamma_g = 5/3, \gamma_c = 4/3$ to be the adiabatic indices of the gas and CR.  `Upstream' means the fluid state at $x=-\infty$, `downstream' means the post-subshock fluid state if there is a subshock or $x=+\infty$ if there is not.

The non-conservative form of the CR subsystem leads to the presence of derivatives in equation \ref{eqn:cr_energy} and \ref{eqn:cr_flux}. This implies that we cannot simply use conservation laws to determine jump conditions, but must solve for the detailed structure of the front. In particular, we must solve ODEs. For this to be possible, the CR variables $P_c,F_c$ (unlike the gas variables) must be continuous across the front. Physically, the smoothness of $P_c,F_c$ across the shock is guaranteed by the large mean free path of CRs, $\lambda \sim r_{\rm c}/(\delta B/B)^{2} \gg \lambda_{i}$, where $r_{g}$ is the CR gyroradius and $\lambda_i$ is the ion mean free path; the (much smaller) thermal ion mean free path sets the characteristic thickness of any gas shock discontinuity. Mathematically, the smooth solutions are guaranteed by the diffusion term in the above equations; we just need to resolve the diffusion length $l_{\rm D} \sim \kappa/c_{\rm s}$. Note that if $P_{\rm c}$ were discontinuous, similar to $P_g$, then equation \ref{eqn:cr_flux} would imply an infinite CR flux $F_c$. 
 
\subsection{Shock Structure and Solution Method} \label{subsec:method}

\subsubsection{Previous Solution: Uni-directional Streaming}

Before solving the above equations, we describe the overall features of the shock. CR acceleration implies that $P_c$ is higher in downstream gas. However, downstream CRs can diffuse upstream and affect the flow. The CR precursor significantly affects fluid flow and decelerates incoming gas, from being supersonic with respect to the overall acoustic speed of the plasma (which includes both gas and CR contributions to gas pressure; $c_{\rm s,tot}^{2} \approx \dv*{(p_g + p_c)}{\rho}$) to subsonic with respect to $c_{\rm s,tot}$. There are two possibilities: (i) in a CR dominated shock, the postshock CR pressure absorbs a significant fraction of the incoming ram pressure. In this case, the `shock' simply consists of a smooth deceleration and compression; all fluid variables are continuous. After the compression, the flow is still supersonic with respect to the {\rm gas} sound speed. (ii) The gas must absorb a significant fraction of incoming ram pressure, an amount which is inconsistent with just adiabatic compression. This implies a discontinuous gas subshock {\it in the gas variables only}, and a jump in gas entropy. The subshock renders the flow subsonic with respect to the gas sound speed. The effect of CR streaming is transfer energy from CRs to the gas in the precursor, preheating the gas and thus increase the importance of gas decelerating the flow, thus increasing the strength of the subshock.  

\emph{The smooth precursor}. The gas is adiabatically compressed. The gas velocity decreases from mass conservation while the gas and CR pressures increase. For a shock propagating in the $-x$ direction, $\nabla P_c > 0$ in the precursor and CR streams towards the upstream ($v_s = -v_A$). The net motion of CR is still towards the downstream as the gas advects faster than $v_A$ ($\mathcal{M}_A \gg 1$). In this region, one can safely take derivatives of the fluid variables, and as shown by VDM84, integrate eqn.\ref{eqn:mass_conv} to \ref{eqn:cr_flux} to yield the `wave adiabat'

\begin{gather}
    \qty{1 + \frac{\mathcal{M}_A}{\gamma_g - 1}}^{2\gamma_g}\qty{P_g + \frac{\qty(\gamma_g - 1)B^2\qty(2\gamma_g \mathcal{M}_A
    + 1 - \gamma_g)}{\gamma_g\qty(2\gamma_g + 1)}} = \text{const}, \label{eqn:adiabat}
\end{gather}

where $\mathcal{M}_A\equiv v/v_A$ is the Alfvenic Mach number. The `wave adiabat' is an additional conserved quantity which relates gas pressure to density. It reduces to the gas entropy $P_g\rho^{-\gamma_g}$ for $\beta\gg \mathcal{M}_A\gg 1$ -- i.e., when $v_{\rm A} \cdot \nabla P_{\rm c}$ is small and there is little energy exchange between CRs and gas, the gas compresses adiabatically. On the other hand, in the limit $\mathcal{M}_A\gg\beta\sim 1$, eqn.\ref{eqn:adiabat} reduces to $\rho = \text{const}$: the gas is incompressible at strong and magnetically significant shocks, due to intense CR heating of the thermal plasma. 

Since we have 4 conserved quantities for 5 variables, only a first order differential equation governing the shock precursor is required to close the system. The precursor equation, expressed in terms of the inverse compression ratio $y=\rho_1/\rho$ (subscript 1 denoting upstream), is \citep{volk84}: 

\begin{equation}
    \dv{y}{x} = \frac{\qty(1-y)N(y)}{\qty(\kappa/v_1)D(y)}, \label{eqn:structure}
\end{equation}

where $N\qty(y)$ and $D\qty(y)$ are given by eqn.24 and 25 in VDM84; we list them here for completeness: 

\begin{gather}
    N\qty(y) = \frac{\qty(\gamma_c + 1)}{2}\qty(y - \frac{\gamma_c - 1}{\gamma_c + 1}) \nonumber \\ \quad - \frac{\gamma_c}{\gamma_g \mathcal{M}^2_{s1}}\left\{ 1 + \delta - \frac{\gamma_g - \gamma_c}{\gamma_c\qty(\gamma_g - 1)}\frac{1 - \bar{P}y}{1 - y} \right\} \nonumber \\ \quad - \frac{\gamma_c}{\mathcal{M}_{A1}}\left\{ y^{1/2} - \frac{1}{\gamma_c \mathcal{M}^2_{c1}\qty(1 + y^{1/2})} + \frac{y^{1/2}\qty(1 - \bar{P})}{\gamma_g \mathcal{M}^2_{s1}\qty(1 - y)}\right\}, \label{eqn:Nold} 
\end{gather}

\begin{gather}
    D\qty(y) = \flatfrac{\qty(\frac{\bar{P}\qty(y)}{y \mathcal{M}_{s1}^2} - 1)}{\qty(1 + \frac{\qty(y - 1)}{\mathcal{M}_{A1} y^{1/2}})} \label{eqn:D}
\end{gather}

where $\delta = P_{c1}/P_{g1}$, $\mathcal{M}_{s1} = v_1/c_{s1}$, $c^2_{s1} = \gamma_g P_{g1}/\rho_1$, $\mathcal{M}_{c1} = v_1/c_{c1}$, $c^2_c = \gamma_c P_{c1}/\rho_1$ and $\bar{P} = P_g/P_{g1}$.

\emph{The subshock}. The subshock is characterized by a set of jump conditions. CR diffusion ensures that only the gas variables jump discontinuously while the CR pressure and flux must be continuous. The jump conditions are therefore:

\begin{gather}
    \qty[\rho v] = 0 \label{eqn:mass_jump} \\
    \qty[\rho v^2 + P_g] = 0 \label{eqn:mom_jump} \\
    \qty[\rho v \qty(\frac{1}{2}v^2 + \frac{\gamma_g}{\gamma_g - 1}\frac{P_g}{\rho})] = 0 \label{eqn:eng_jump} \\
    \qty[P_c] = \qty[F_c] = 0 \label{eqn:cr_cont}
\end{gather}

From the jump conditions, one can derive the relation: 

\begin{equation}
    \gamma_g\langle P_g \rangle = J\langle v \rangle \label{eqn:jump}
\end{equation}

where $\langle\cdot\rangle$ denotes the arithmetic mean of the enclosed quantity just before and after the jump and $J=\rho v$ is the conserved mass flux.

What is the criterion for a gas subshock? It occurs when the compression ratio $y$ is discontinuous, i.e. in equation \ref{eqn:structure}, $\dv*{y}{x} \rightarrow \infty$ when $D\qty(y) = 0$ (it can be shown that $N\qty(y)$ is finite, even in the limit $y\rightarrow 1$). From equation \ref{eqn:D}, we see that this happens when:

\begin{equation} 
v^{2} = \frac{\gamma_g P_g}{\rho} = c_{\rm s}^{2}, \label{eqn:sonic} 
\end{equation}

i.e. the flow hits a sonic point with respect to the {\it gas} sound speed. We see that this is equivalent to equation \ref{eqn:jump} derived from the jump conditions. Since fluid variables are discontinuous at a shock, the sonic point is defined in terms of the average of pre-shock and post-shock quantities. The upstream flow is of course supersonic; if the downstream flow is still supersonic with respect to the gas sound speed, then there is no sonic transition and no subshock. We still refer to the entire compressive structure as a `shock', since the fluid decelerates from $\mathcal{M} > 1$ to $\mathcal{M} < 1$ with respect to the {\it total} sound speed\footnote{Note that this differs from simply summing the gas and CR pressure to get the total pressure in an adiabatic medium, because energy is transferred between the gas and CRs.}, given by VDM84:

\begin{gather}
    v_p^2 = c_s^2 + c_c^2\frac{\qty(v - v_A/2)\qty(v + \qty(\gamma_g - 1) v_A)}{v\qty(v - v_A)}, \label{eqn:vp}
\end{gather}

where $c_c^2 = \gamma_{c} P_{c}/\rho$. However, if the downstream flow is subsonic, then equation \ref{eqn:structure} becomes singular at the sonic point and a subshock occurs.  

The sonic point is where $P_c$ is maximized. Physically, this is because at the subshock, the kinetic energy of the flow goes into the gas component rather than the CR component: $P_g$ ungoes a discontinuous increase at the subshock, while $P_c$ is unchanged (continuous) across the subshock. After the subshock, one goes directly to the downstream state where all fluid variables are constant. One can also see this by differentiating equation \ref{eqn:energy_conv} and using equation \ref{eqn:cr_energy} to obtain: 

\begin{equation}
    \qty(\rho v^2 - \gamma_g P_g)\dv{\rho}{x} = \rho\dv{P_c}{x} \label{eqn:max_P_c} 
\end{equation}

i.e. as one approaches the sonic point where the term in brackets vanishes, $\grad P_c \rightarrow 0$ and $P_c$ is maximized. Note that if the solution were to remain continuous and differentiable, then $\grad P_c$ would change sign, implying a non-monotonic precursor profile, which is unphysical in the presence of diffusion. One can also see that at a sonic point, $\dv*{y}{x}$ would change sign since $D(y)$ changes sign (see equation \ref{eqn:D}), again implying a non monotonic profile. However, if a subshock takes place at the sonic point, derivatives involving the gas diverge (in particular, $\dv*{\rho}{x} \rightarrow \infty$) and so all equations involving derivatives (including equation \ref{eqn:structure} and \ref{eqn:max_P_c}) are no longer valid.  

\begin{figure}
    \centering
    \includegraphics{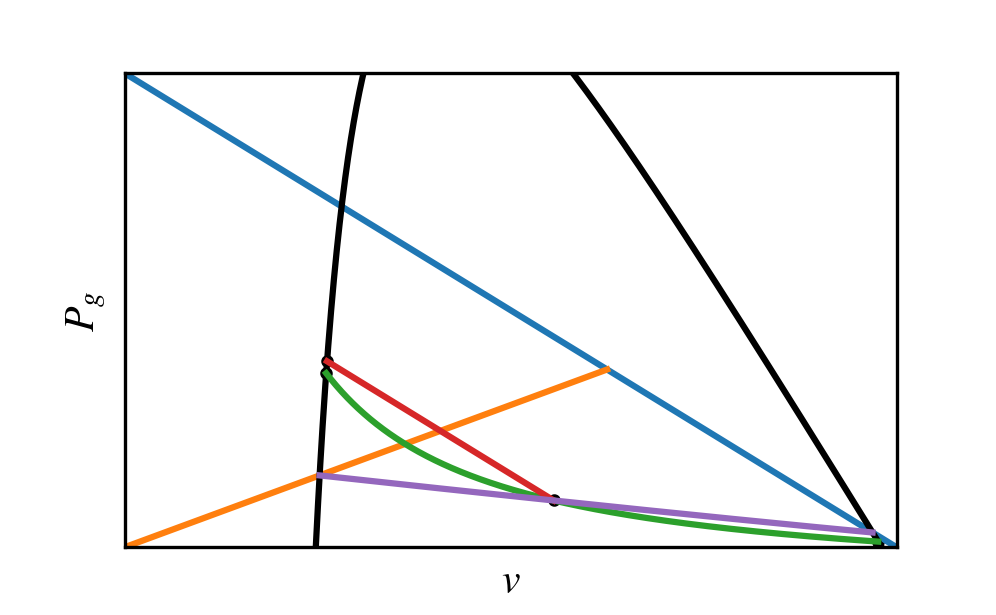} \\
    \includegraphics{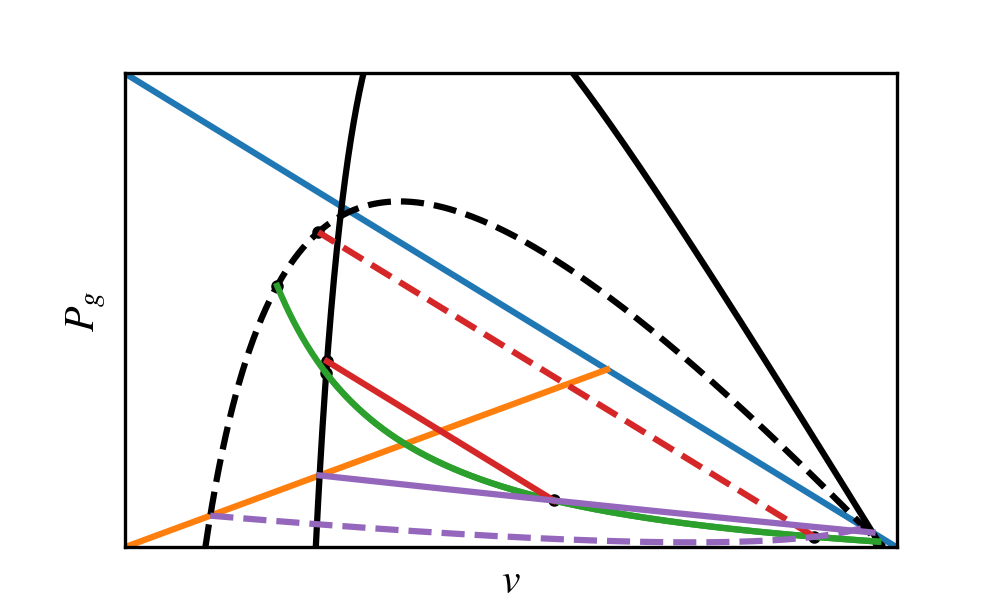}
    \caption{\emph{Top}: Typical $P_g$ against $v$ diagram without bi-directional streaming. Each colored curve represents a $P_g-v$ relation given a condition, as described in \citet{drury81}. Blue - $P_c = 0$; orange - $\gamma_g P_g = J v$; black - the Hugoniot, $N\qty(y) = 0$; red - jump in $\qty(P_g, v)$ satisfying the subshock jump conditions \ref{eqn:mass_jump} - \ref{eqn:cr_cont}; green - the wave adiabat, eqn.\ref{eqn:adiabat}; purple - the reflected Hugoniot. \emph{Bottom}: Same as the top but with bi-directional streaming, which leads to the dotted lines. The dotted black line expresses the $P_g-v$ relation for $\tilde{N}\qty(y) = 0$. Construction of the dotted red and purple line follows that of the solid red and purple line. For reference, $\mathcal{M} = 5, Q = 0.5, \beta=1$ for these two plots.}
    \label{fig:sol_diag}
\end{figure}

\emph{Solution method}. In the standard treatment by VDM84, for a given upstream, the downstream can be found by a modification of the procedure described in \citet{drury81} (hereafter DV81). The solution procedure can be expressed graphically, as in the top panel of Fig.\ref{fig:sol_diag}, which shows a $P_g$ against $v$ diagram. Each curve on the diagram describes a constraint characteristic to the shock structure: 
\begin{itemize} 
\item{{\it $P_c=0$ (blue curve).} Pressure must be positive. The plot shows $P_g > 0$ only; another obvious constraint is $P_c > 0$. Thus, all valid solutions must lie below the blue line, which shows $P_c=0$ (obtained from equation \ref{eqn:mom_conv}). Lines parallel to this line correspond to $P_c=\mathrm{const}$, which we will use shortly.}
\item{{\it Hugoniot (black curve).} The black curve references eqn.\ref{eqn:structure}, showing where $N\qty(y) = 0$, or equivalently where the gradients of the fluid variables are zero. This corresponds to far upstream and downstream. It is called the Hugoniot. For a given upstream, the Hugoniot encompasses possible downstream states.} 
\item{{\it Wave adiabat (green curve).} The wave adiabat, given by equation \ref{eqn:adiabat}, is set by upstream conditions and conserved throughout the precursor. The initial intersection of the wave adiabat and the Hugoniot at the far right gives the upstream state; the subsequent intersection at the left gives the downstream state if there is no subshock. The ordering of these states is unambiguous, since the shock decelerates the flow.} 
\item{{\it Sonic Boundary (orange line).} The orange line shows the sonic condition given by equation \ref{eqn:sonic}. If the wave adiabat does not cross this boundary before reaching the Hugoniot, then it never undergoes a sonic transition and there is no subshock. The structure of the shock can then be read off graphically by following the wave adiabat from the upstream to the downstream state. On the other hand, if it crosses this line, then the gas will shock.}
{\item {\it Reflected Hugoniot (purple line).} If the gas undergoes a sub-shock, how do we proceed? Since $P_c$ is continuous, $[P_c]=[\rho v^{2} + P_{g}]=0$ across the subshock, the jump in fluid variables must be parallel to the $P_c=0$ (blue) line. In addition, from equation \ref{eqn:jump}, the sonic boundary (orange line) must bisect this line, since the sonic boundary gives the relationship between the {\it mean} of the pre-shock and post-shock pressure and velocities. From these facts, we can construct a `reflected Hugoniot' (purple curve), which is the locus of points traced out by lines parallel to the blue $P_c=0$ line, which start at the Hugoniot (black) and are bisected by the sonic boundary. The reflected Hugoniot shows all the possible pre-subshock states connected to the downstream by the subshock jump conditions eqn.\ref{eqn:mass_jump}-\ref{eqn:cr_cont}. The intersection of the wave adiabat (green) and reflected Hugoniot (purple) therefore gives the pre-subshock state.}
{\item {\it Subshock Jump (red line).} Now that we have identified the pre-subshock state, we insert the subshock jump (red line), which as discussed must be parallel to the $P_c=0$ (red) line. The intersection of the subshock jump (red line) with the Hugoniot (black line) gives the post subshock (and final downstream) state.} 
\end{itemize} 

In summary, the solution procedure is: follow the wave adiabat (green) in the direction of decreasing $v$ until it intersects with the reflected Hugoniot (purple), then follow the subshock jump (red) directly to the downstream. In the absence of a subshock, possible if the wave adiabat does not cross the sonic boundary (orange), the downstream is simply given by the intersection of the Hugoniot and the wave adiabat. Such smooth transitions can occur if the shock is CR dominated. 

The solution can be parametrized by: 

\begin{equation}
    \mathcal{M} = \frac{v_1}{v_{p1}},\quad Q = \frac{P_{c1}}{P_{g1} + P_{c1}},\quad \beta = \frac{8\pi P_{g1}}{B^2}, \label{eqn:m_n_beta}
\end{equation}

where $v_{p}$ is given by eqn.\ref{eqn:vp} here. The shock Mach number $\mathcal{M}$ is not to be confused with the Alfvenic Mach number $\mathcal{M}_A$, the sonic Mach number $\mathcal{M}_s$, or the CR acoustic Mach number $\mathcal{M}_c$. $Q$ is the upstream non-thermal fraction of the total pressure. $\beta$ is the familiar plasma beta.

\subsubsection{New Solution: Bi-directional Streaming}

\begin{figure}
    \centering
    \includegraphics[width=0.45\textwidth]{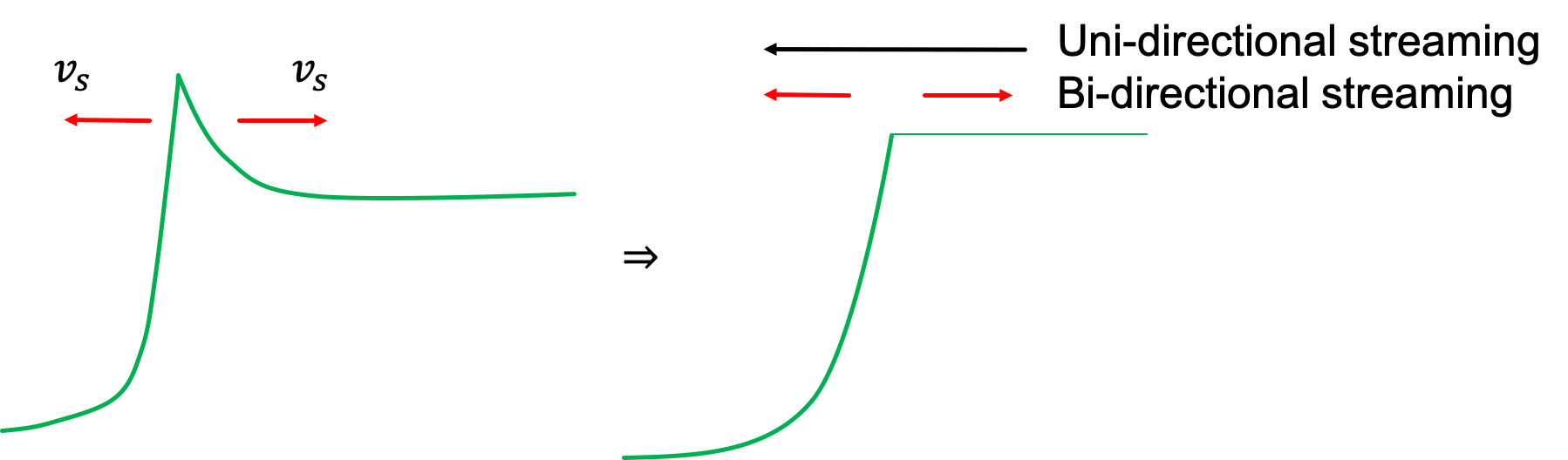}
    \caption{Conceptual plot of $P_c$ leading to bi-directional streaming. The direction of streaming assumed in VDM84 is added for comparison.}
    \label{fig:spike}
\end{figure}

The aforementioned solution method assumes the direction of CR streaming is the same throughout the shock profile, i.e. towards the upstream. However, post-subshock CR can stream towards the downstream too. At the early stages of shock formation, strong compression at the subshock can cause the CR pressure to overshoot, forming a small spike from which CR stream away in opposite directions (Fig.\ref{fig:spike}). This is entirely analogous to the `Zeldovich spike' \citep{zeldovich67} which occurs in radiative shocks. The spike is a non-equilibrium state which slowly flattens as CRs stream out. However, it sets up a shock structure where downstream CRs stream {\it away} from the shock, rather than towards it, as VDM84 assumed. Note that the downstream CR profile is almost flat ($P_c\rightarrow\rm const$), so the direction of streaming is set by small changes in the CR profile at the shock. 

To capture this new solution graphically, a new Hugoniot curve has to be added (see bottom panel of fig.\ref{fig:sol_diag}). This new Hugoniot is derived by setting $\tilde{N}\qty(y) = 0$, where $\tilde{N}\qty(y)$ is the function $N\qty(y)$ with the signs in front of $\mathcal{M}_A$ flipped,

\begin{gather}
    \tilde{N}\qty(y) = \frac{\qty(\gamma_c + 1)}{2}\qty(y - \frac{\gamma_c - 1}{\gamma_c + 1}) \nonumber \\ \quad - \frac{\gamma_c}{\gamma_g \mathcal{M}^2_{s1}}\left\{ 1 + \delta - \frac{\gamma_g - \gamma_c}{\gamma_c\qty(\gamma_g - 1)}\frac{1 - \bar{P}y}{1 - y} \right\} \nonumber \\ \quad + \frac{\gamma_c}{\mathcal{M}_{A1}}\left\{ y^{1/2} - \frac{1}{\gamma_c \mathcal{M}^2_{c1}\qty(1 + y^{1/2})} + \frac{y^{1/2}\qty(1 - \bar{P})}{\gamma_g \mathcal{M}^2_{s1}\qty(1 - y)}\right\}, \label{eqn:Nnew} 
\end{gather}

The standard Hugoniot (solid black line in fig.\ref{fig:sol_diag}) shows possible downstream solutions for which $v_s = -v_A$, where post-shock CR streams toward the shock. With the sign flip, the new Hugoniot (dotted black line) shows possible downstream solutions for which $v_s = v_A$, and the post-shock CR stream away from the shock. The switch in direction of downstream CRs changes not just the magnitude of the subshock, but also where it occurs. One can see it is not possible to jump, from the standard location where the subshock occurs (intersection between the solid green and purple lines), to the new Hugoniot while satisfying the subshock jump conditions. The sonic boundary (orange) would no longer bisect the line connecting pre-shock and post-shock states. To determine when the subshock occurs, a new reflected Hugoniot (dotted purple line) has to be calculated, in a similar manner as in the standard treatment.

Fluid flow in the precursor conserves the same wave adiabat as before since CR still streams towards the upstream. Therefore precursor fluid states continue to trace the same green curve. The subshock occurs at intersections between the new reflected Hugoniot and the wave adiabat, which brings the fluid directly to the downstream.

\emph{Existence of the new solution}. In the case of uni-directional streaming, the shock profile can be smooth (i.e. no subshock) if the wave adiabat does not cross the sonic boundary. This happens when the upstream $P_c$ is sufficiently high. However, the new solution always requires a subshock. Bi-directional streaming can only occur if there is a maximum in $P_c$, at which $\nabla P_c=0$. As previously discussed (see equation \ref{eqn:max_P_c}), unless $\dv*{\rho}{x} = 0$, a maximum in $P_c$ is equivalent to a sonic point in the gas, and thus a subshock must occur, which brings the fluid to its downstream state without further relaxation. Otherwise, the profile will be non-monotonic. This means that in CR dominated regimes, the new solution may cease to exist because the subshock has been smoothed out.

\begin{figure}
    \centering
    \includegraphics{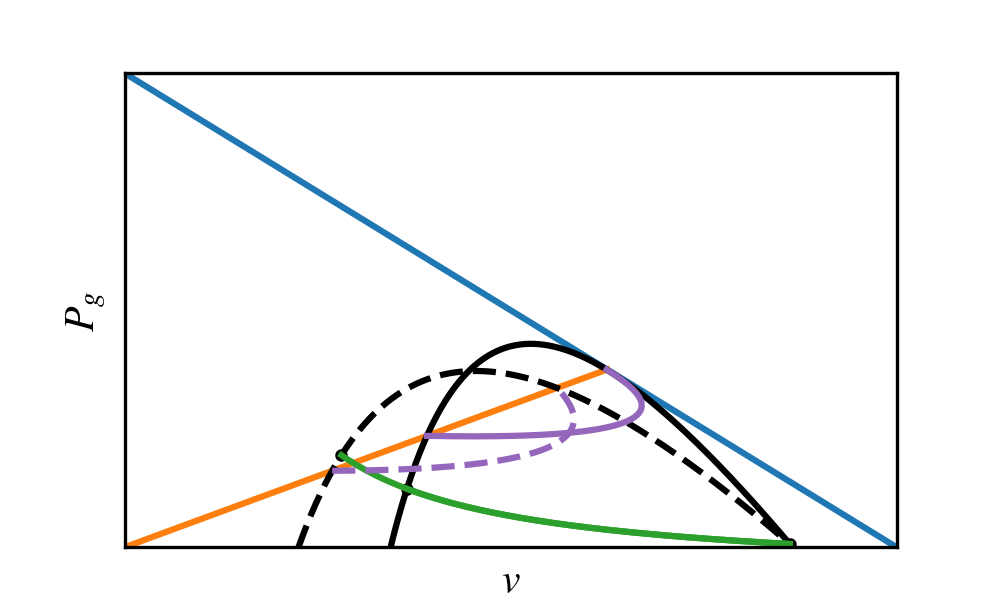} \\
    \includegraphics{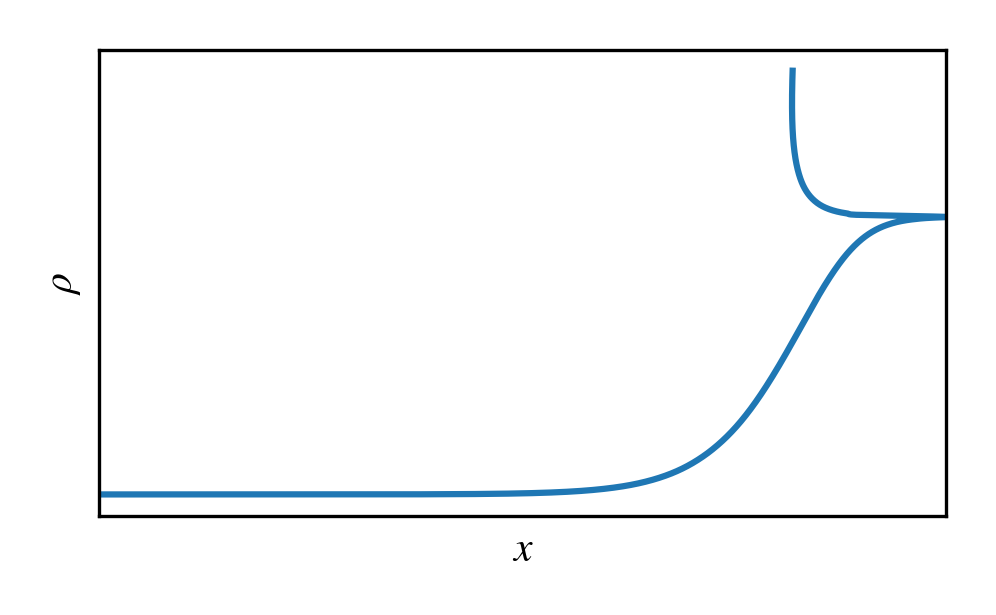}
    \caption{\emph{Top}: $P_g-v$ diagram for $\mathcal{M} = 2, Q = 0.95, \beta = 1$. The color of the curves mean the same as in fig.\ref{fig:sol_diag}. \emph{Bottom}: Density plot of the new solution shows that it is non-monotonic.}
    \label{fig:sol_diag_no_new}
\end{figure}

Fig.\ref{fig:sol_diag_no_new} shows an example where a new solution is not allowed for the above reasons. The wave adiabat (green) does not cross the sonic boundary before intersecting with the standard Hugoniot (black). Thus, the standard solution involves a smooth transition, with no sub-shock. The wave adiabat (green) does also intersect with the new reflected Hugoniot (dotted purple), but only after crossing the standard Hugoniot (solid black), where $\dv*{P_c}{x} = 0$. Continuing after this would imply a change in sign for $\dv*{P_c}{x}$ and other fluid derivatives, i.e. a non-monotonic profile. This solution therefore has to be rejected. 

\subsection{Solution Structure} \label{subsec:sol_struct}

\begin{figure}
    \centering
    \includegraphics{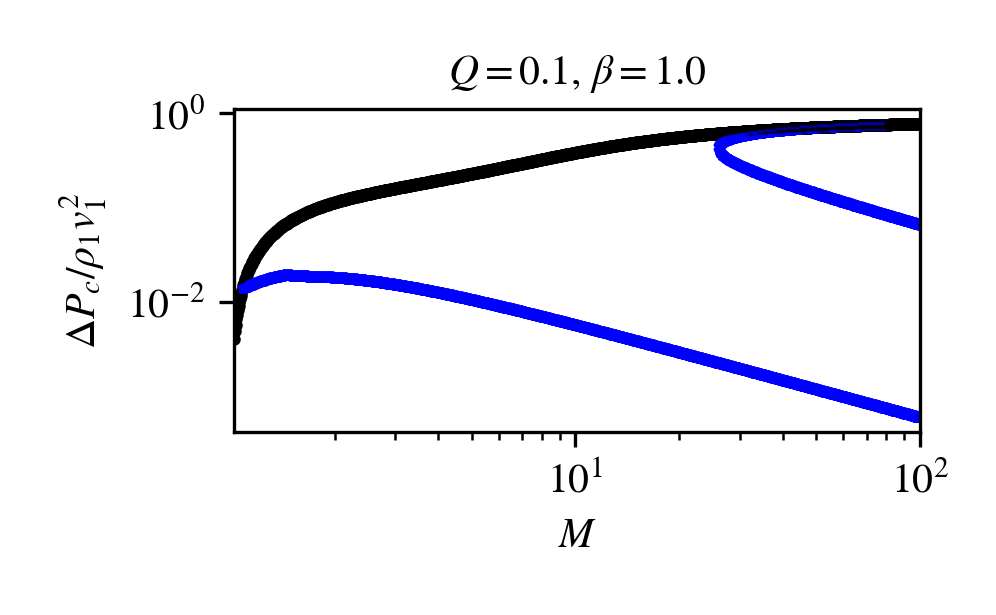} \\
    \includegraphics{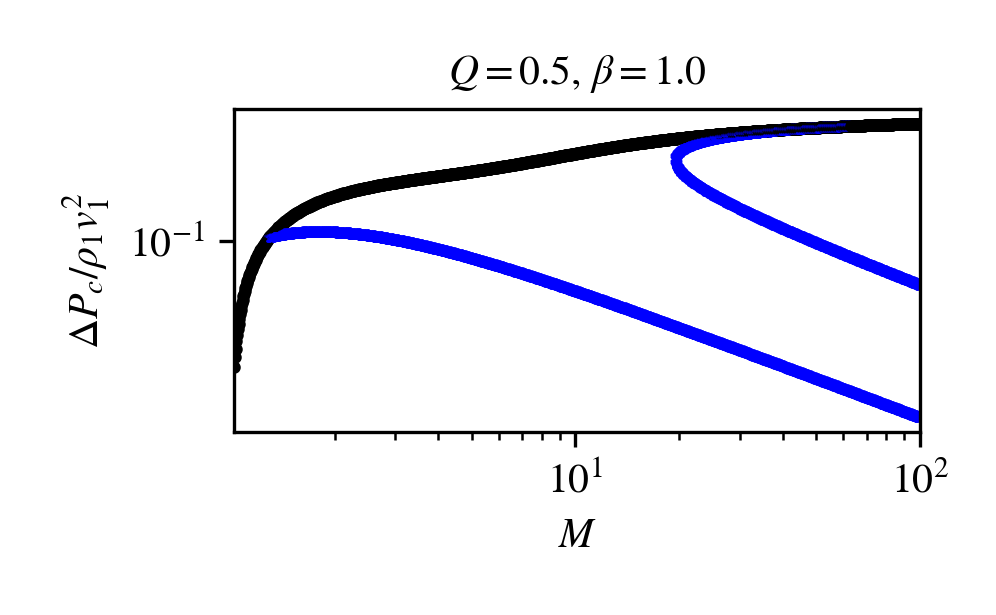}
    \caption{Acceleration efficiency against mach number $\mathcal{M}$ for $Q = 0.1$ (Top) and $= 0.5$ (Bottom) and $\beta = 1$. The black curve denotes the standard branch while the blue curve denotes the new solution branches (efficient, intermediate and inefficient).}
    \label{fig:mach}
\end{figure}

\begin{figure*}
\centering
\includegraphics{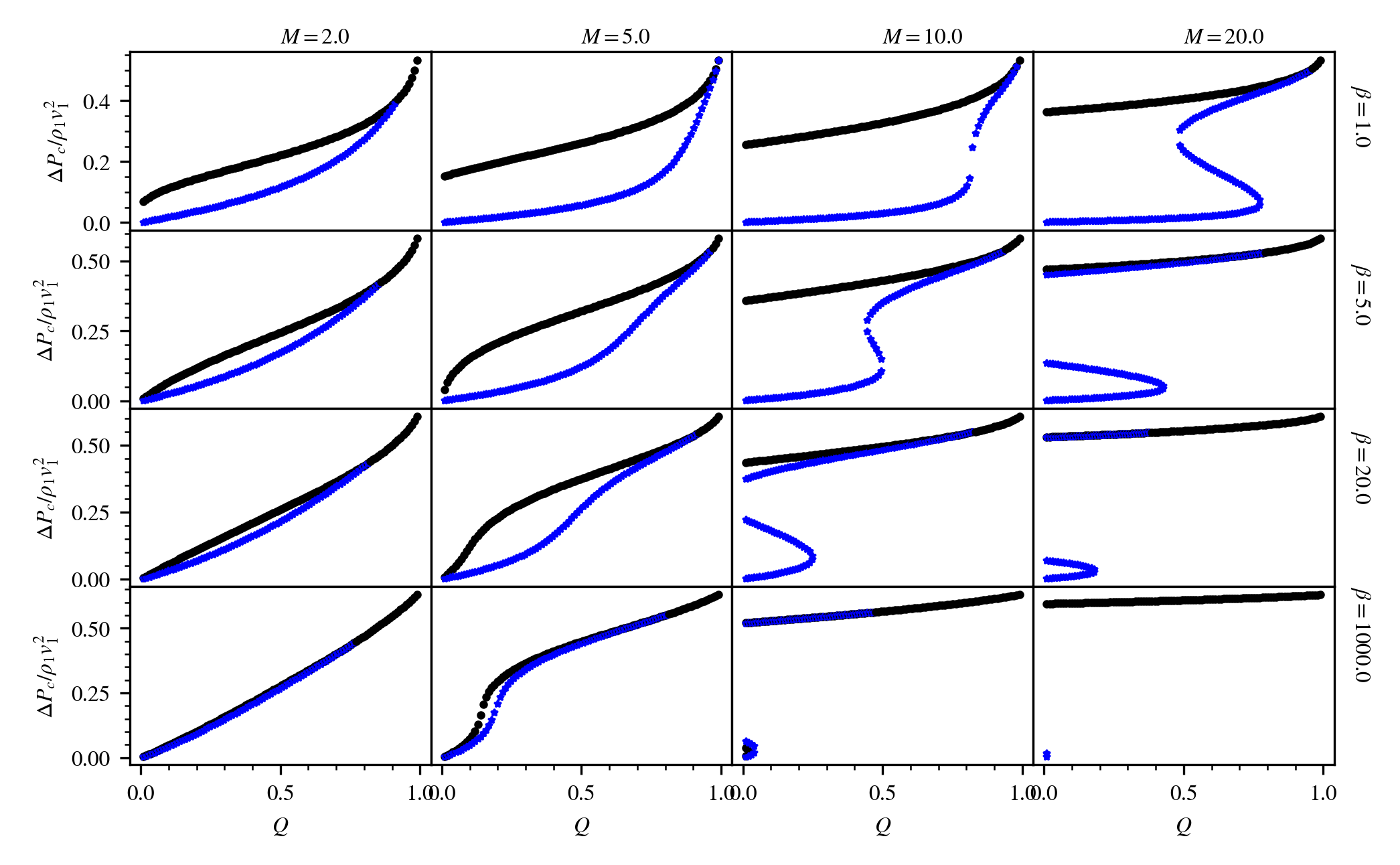}
\caption{Acceleration efficiency against upstream non-thermal fraction ($Q$) for Mach number $\mathcal{M}=2,5,10,15$ and plasma beta $\beta=1,5,20,1000$. The black curve denotes the standard branch while the blue curve denotes the new solution branches (efficient, intermediate and inefficient). In each panel, the efficient branch gradually merges with the standard branch as $Q$ increases. At sufficiently CR dominated regimes (high $Q$), the efficient branch cease to exist by the monotony argument in sec.\ref{subsec:method}.}
\label{fig:sol_struct}
\end{figure*}

\begin{figure*}
    \centering
    \includegraphics{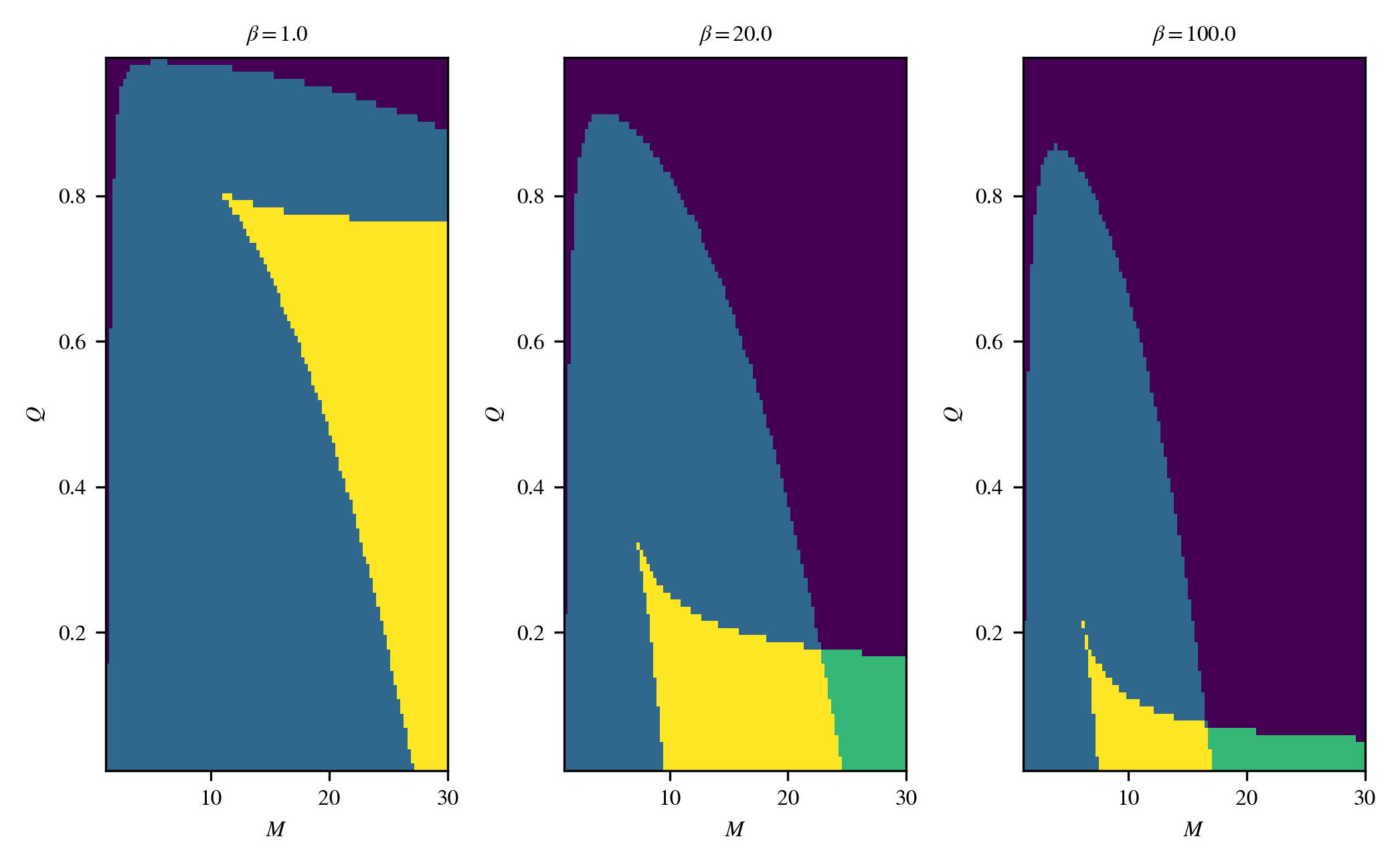}
    \caption{Color plots showing the multiplicity of the new + standard solution for a given $\qty(\mathcal{M}, Q, \beta)$. Purple = 1, blue = 2, green = 3, yellow = 4.}
    \label{fig:sol_number}
\end{figure*}

Fig.\ref{fig:mach} shows the acceleration efficiency, measured by the ratio of the change in CR pressure to the upstream ram pressure ($\qty(P_{c2} - P_{c1})/\rho_1 v_1^2 \equiv \Delta P_c/\rho_1 v_1^2$), against Mach number for upstream non-thermal fraction $Q = 0.1, 0.5$ and $\beta = 1$. Fig.\ref{fig:sol_struct} shows the acceleration efficiency against $Q$ for a sample of Mach number and plasma beta. In these two figures, two different solutions emerge, corresponding to uni-directional (black curves) or bi-directional streaming (blue curves). At high $\beta$, the two solutions converge since the contribution of streaming is small in that limit, so it does not matter which way the CRs stream. In magnetically significant regimes ($\beta\sim 1$), the new branch introduces two main differences: first, the acceleration efficiency is in general lower. For bi-directional streaming, downstream CRs stream away from the subshock, and fewer CRs diffuse to the upstream precursor. In the two-fluid formalism, all CR `acceleration' is essentially compressional (adiabatic) heating. With a smaller precursor, the shock is more hydrodynamic and less compressible (since the $\gamma_{g}=5/3$ gas is less compressible than the $\gamma_c=4/3$ CRs). Lower compression implies less overall less adiabatic heating of the CRs. The difference is small at low Mach numbers ($\mathcal{M}\sim 1-2$) but becomes more apparent as $\mathcal{M}$ increases. At $\mathcal{M}\sim 10$ the acceleration efficiency can drop from $\sim 40-50\%$ for the standard branch to less than $10\%$. However, at moderate Mach numbers, a transition occurs, and that brings us to the second point: the new solution bifurcates into multiple branches. A similar bifurcation occurs for CR shocks without streaming, which is equivalent to our high $\beta$ limit (DV81; \citealt{jones91, donohue93, mond98, becker01, saito13}). This does not happen for uni-directional solutions with streaming. Even so, bifurcation in the no streaming case happens only at very small $Q$, and within an intermediate range of Mach numbers. In contrast, the new bifurcation can occur at high $Q$ (for $\beta\sim 1$, it can occur for equipartition CR energy densities or even in CR dominated regimes) and persists even as $\mathcal{M}$ continue to increase. Fig.\ref{fig:sol_number} shows a summary of the solution multiplicity for $\qty(\mathcal{M}, Q)$ and $\beta=1,20,1000$. Multiple solutions for the new branch are common, particularly for high Mach numbers and when magnetic fields are significant (lower $\beta$). The new branch usually bifurcates into three solutions, and in order of increasing acceleration efficiency we shall call them the inefficient, intermediate and efficient branch. 

The uni-directional solution poses a difficulty at low $\beta$: a significant downstream non-thermal fraction exists even as $Q\rightarrow 0$. This solution has been argued to be physically unrealistic \citep{malkov97}. It is unclear physically how, without injection, one can accelerate particles without an existing CR population. By contrast, the bi-directional solution has at least one branch where $\Delta P_c \rightarrow 0$ as $Q \rightarrow 0$. 

The behaviour of different branches of solution are markedly different. The `inefficient' branch corresponds to the test particle limit where CRs have nearly no effect on the shock structure. The downstream fluid is gas dominated and the shock appears hydrodynamic, giving a compression ratio $\sim 4$ at high Mach number. At equipartition ($Q\sim 0.5, \beta\sim 1$), the typical acceleration efficiency $\lesssim 10\%$ for Mach numbers below $10$ and {\it decreases} with increasing Mach number (see fig.\ref{fig:mach}). At low Mach numbers, the acceleration efficiency of this branch appears consistent with PIC/hybrid simulations \citep{caprioli14a}, which found an efficiency of $\lesssim 10\%$ for $\mathcal{M} < 10$. It is however at odds with the behavior at very high Mach numbers $\mathcal{M} > 10$, for which PIC/hybrid simulations show an increase in efficiency. We shall see that if we include thermal injection into DSA (\S\ref{sec:thermal}), this decline in efficiency at high Mach number goes away. 

The efficient branch is strongly CR modified. Having a smaller adiabatic index $\gamma_c = 4/3$, the fluid is more compressible, so the compression ratio $\sim 7$ at high Mach number. This leads to much higher adiabatic heating of the CRs. At equipartition, this branch emerge at Mach numbers higher than $\sim 12$ and has a typical efficiency of $\gtrsim 60\%$. The acceleration efficiency continues to increase with Mach number such that at Mach number of a few tens and above the subshock is smoothed out by the dominating CR population and the efficient branch merges with the standard branch. In the following, we shall often refer to the efficient branch and standard solution collectively as the efficient/standard branch due to their similarity in acceleration efficiency. An acceleration efficiency of this order has been found in previous works, both analytically (\citealt{caprioli08}, in a two-fluid model; \citealt{caprioli09}, in a kinetic description; both works include magnetic field amplification) and in simulations (\citealt{ellison84}, in a Monte Carlo approach).

The downstream $P_c$ can be different by decades across branches of solution, so knowing which one nature selects is important. We seek to answer this with simulation. In the following sections, we will demonstrate numerically that the standard and new solutions are all valid steady state shock profiles, but the intermediate branch of the new solution is unstable. We also illustrate, with different initial setups, how various branches can be captured. They turn out to be sensitive to local upstream conditions, but generically, the inefficient branch of the new solution is the one most likely to be realized in realistic settings. 

\section{Simulation} \label{sec:simulation}

\subsection{Code} \label{subsec:code_and_setup}

The following simulations were performed with Athena++ \citep{stone20}, an Eulerian grid based MHD code using a directionally unsplit, high order Godunov scheme with the constrained transport (CT) technique. CR streaming was implemented with the two moment method introduced by \citet{jiang18}. This code solves equation 9 in \citet{jiang18}, which reduces to our equation \ref{eqn:two_moment} in 1D (where the B-field is constant and parallel to the shock normal). Unless otherwise specified, a 1D Cartesian grid is used and the magnetic field points in the $+x$ direction.

\subsection{Setup 1: Imposed Shock Profile} \label{subsec:impose}

We begin by verifying the analytic standard and new solutions, by imposing the steady-state analytic  profiles as initial conditions and verifying that they are time-steady in the code. For a given upstream state, the downstream state can be determined by the method described in \S\ref{subsec:method}. The shock profiles can be calculated from eqn.\ref{eqn:structure} supplemented by a subshock that brings the fluid to the downstream state. This profile is input into the simulation domain and evolved in time. Since the shock structure depends only on $\mathcal{M}, Q, \beta$, we fix the upstream $\rho = 100, P_g = 1$ in code units, implying an unstream gas sound speed of $c_{\rm s} = 0.13$. We set the reduced speed of light $\tilde{c} = 100$\footnote{The reduced speed of light $\tilde{c}$ is a free simulation parameter governing the CR free stream speed in the decoupled limit. It should be much greater than other characteristic speeds. See \citep{jiang18} for a detailed discussion.}. Some simulations were rerun with $\tilde{c}=1000$ with no apparent difference. The diffusion coefficient (which we set to $\kappa=0.1$) has no effect on downstream values, it only sets the shock width. The number of grid cells is $4096$; at this resolution the diffusive length is typically resolved with $n_\mathrm{shock}\equiv \kappa/v_1\Delta x\gtrsim 40$ cells. Previously, $n_\mathrm{shock}\sim 10-20$ was found to be sufficient for convergence \citep{frank94}. Outflow boundary conditions were used on both sides. The result is independent of the boundary conditions as increasing the domain size and imposing the ghost zones yield no difference. Unless specified, the following simulations assume $\beta=1$. CR transport at high $\beta$ is purely diffusive, a limit that has been extensively studied, which we will not investigate in this work. 

\begin{figure*}
    \centering
    \subfigure{\includegraphics{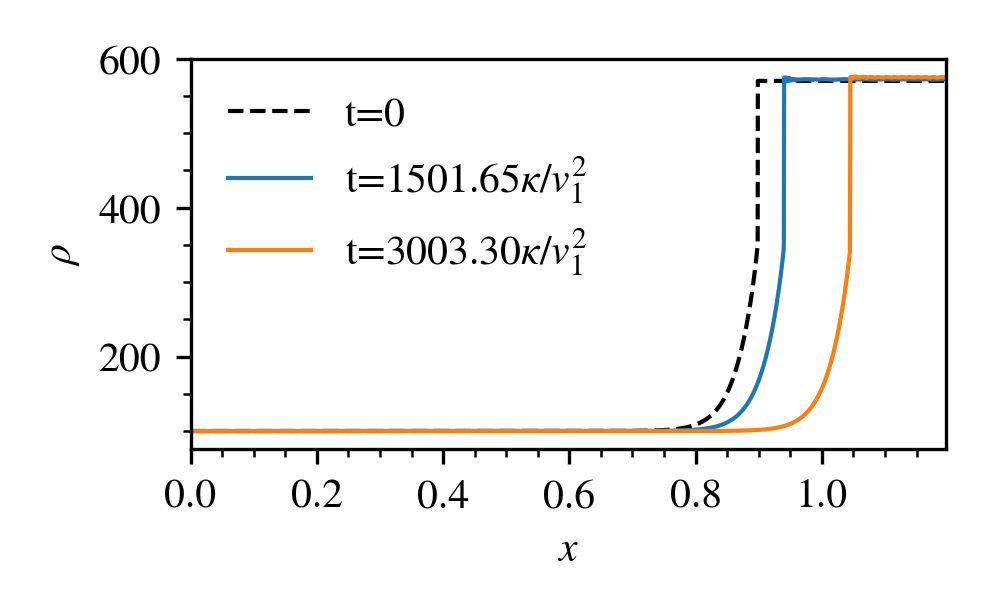}}
    \subfigure{\includegraphics{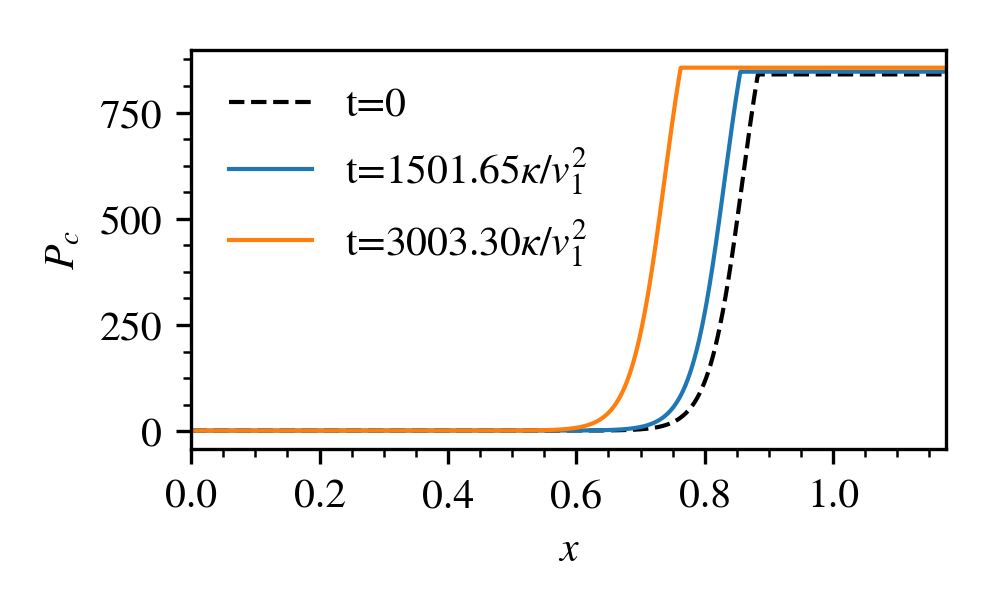}} \\
    \subfigure{\includegraphics{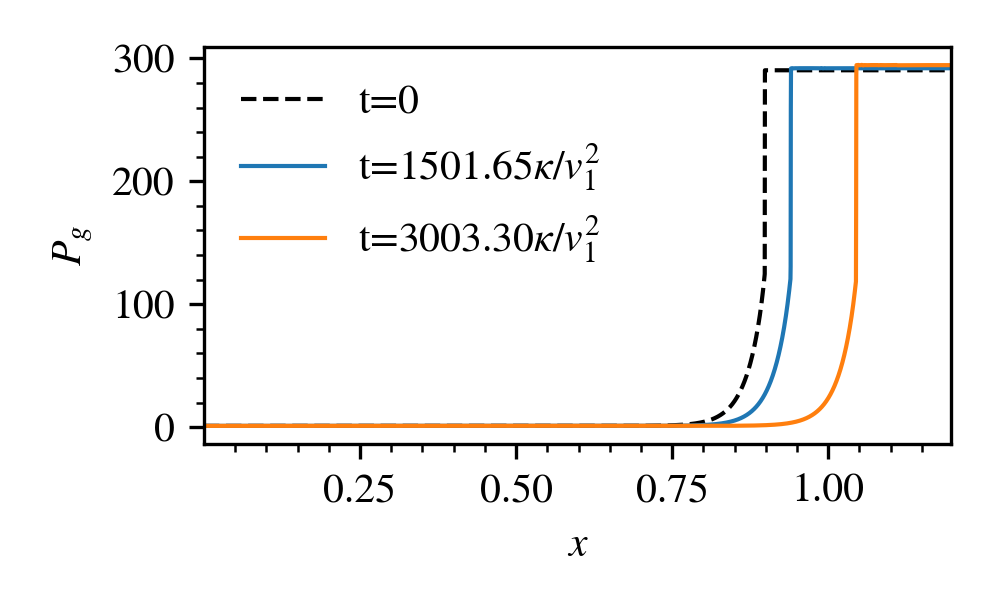}}
    \subfigure{\includegraphics{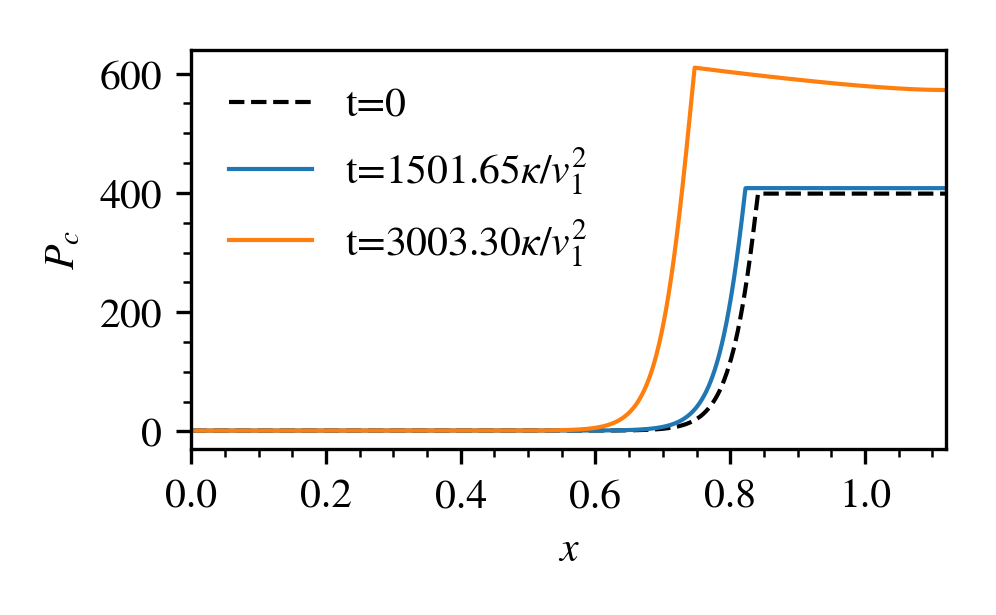}} \\
    \addtocounter{subfigure}{-4}
    \subfigure[]{\includegraphics{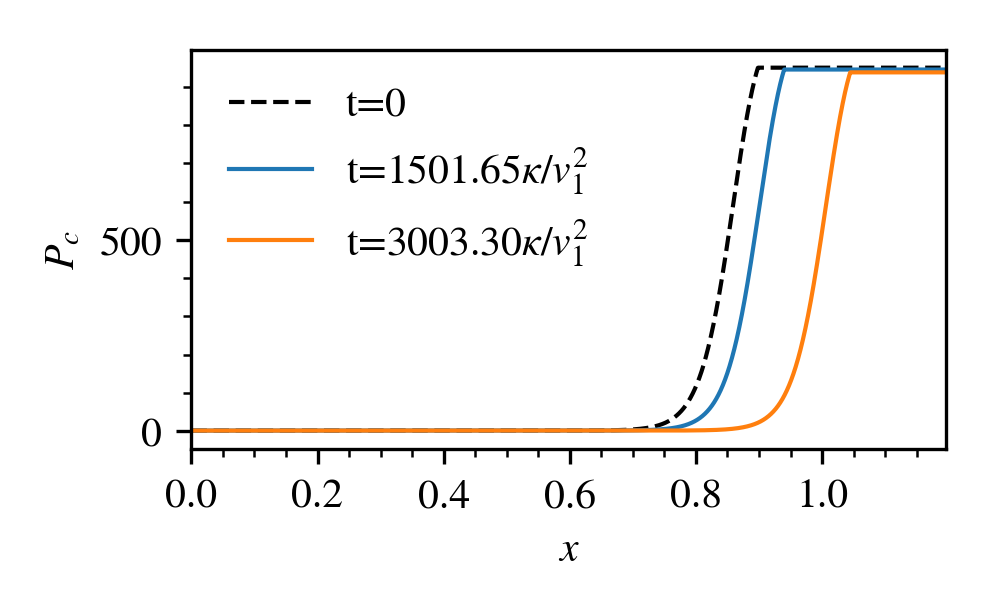}\label{fig:early_std}}
    \subfigure[]{\includegraphics{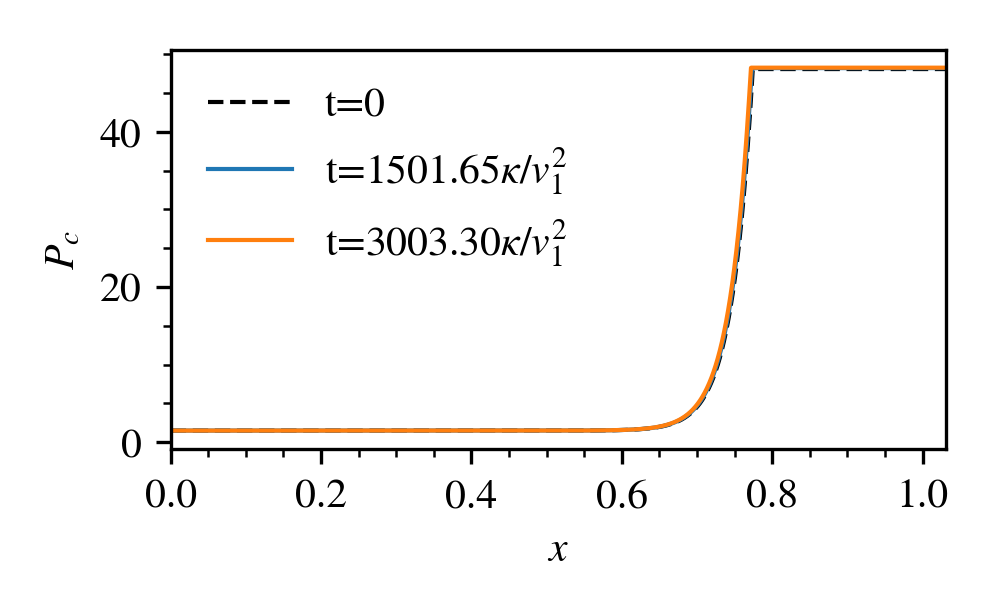}\label{fig:early_pc}}
    \caption{Simulation results for Setup 1. Fig.\ref{fig:early_std}: Shock profiles of the standard branch for $\mathcal{M}=20,Q=0.6,\beta=1$ at three time instances. Fig.\ref{fig:early_pc}: $P_c$ profiles of the new solution branches for the same upstream parameters as (a).\emph{Top}: Efficient branch. \emph{Middle}: Intermediate branch. \emph{Bottom}: Inefficient branch.}
    \label{fig:early}
\end{figure*}

Fig.\ref{fig:early_std} shows the shock profiles of the standard branch while fig.\ref{fig:early_pc} shows the $P_c$ profiles only (for simplicity) of the 3 bi-directional solution branches at $t=0, 1502 t_{\text{diff}}, 3003t_{\text{diff}}$ for $\mathcal{M}=20, Q=0.6, \beta=1$, where $t_{\text{diff}}=\kappa/v_1^2$ is the diffusion timescale. The solutions at $t=1502 t_{\text{diff}}$ are relatively well maintained, with a little numerical shift. Such numerical shifts are expected to equilibrate after $\sim 1000 t_{\text{diff}}$ \citep{kang90}. Behavior of the solutions vary drastically after $\sim 2000 t_{\text{diff}}$, with the intermediate branch diverging exponentially from its original profile. The standard and efficient branch show small spatial shifts, but overall the profile is maintained, with the same acceleration efficiency. The inefficient branch appears the most robust. In general, solution branches with significant downstream CR fraction tend to be the most susceptible to this numerical shift. Such shift originate at the subshock (we do not observe this for smooth shocks). The problem lies with how the direction of the streaming velocity $v_s$ is determined. The direction is determined by $\sign{\nabla P_c}$, which is estimated by a finite difference scheme: $\sign{\nabla P_{c,i}} = \sign{(P_{c,i+1} - P_{c,i-1})/\Delta x}$. The cells at the subshock therefore would still have positive $\nabla P_c$ whereas it should, for bi-directional streaming, be negative. This causes $v_s$ to be positive at the subshock and $F_c$ (equation \ref{eqn:cr_flux}) to overshoot slightly, hence causing the shift in $P_c$ profile. Since $F_c$ scales linearly with $P_c$, the overshoot is larger for more CR dominated downstream. The inefficient branch, which has the lowest downstream CR fraction, is therefore the least affected. Another possible source of shift comes from the finite coupling time for $F_c$ to attain its steady state value (eqn.\ref{eqn:cr_flux}), given roughly by $t_\mathrm{coup} = 1/\sigma_c \tilde{c}^2$. Across the subshock, $\sigma_c$ drops abruptly, leading to a rise in the coupling time. Deviations of $F_c$ from the steady state expression (eqn.\ref{eqn:cr_flux}) causes the tiny discrepancy seen.

The intermediate branch has a profile which does not just translate spatially; it also clearly evolves. Furthermore, in the example above, the acceleration efficiency of the intermediate branch diverges with time and evolves to the standard/efficient branch efficiencies, while the other branches remain close to their initial values. Thus, the intermediate branch is unstable. The same multiplicity (3) of solutions appears in standard solutions with diffusion only, and the intermediate branch is also unstable in this case. \citet{mond98} suggested that this divergent behavior is a consequence of a corrugational instability. Nevertheless, along with \citet{donohue93}, \citet{saito13}, we have found that the intermediate branch is unstable without invoking a corrugation mode (since our simulations are 1D). It is also unlikely to be due to the acoustic instability \citep{drury84, dorfi85, zank85, drury86, kang92, wanger06}, triggered at the shock precursor by phase shifts between the acoustic disturbances in the gas and CR components due to CR diffusivity: the typical growth time (i.e. e-folding time) of the acoustic instability is $t_\mathrm{grow}\sim\kappa/c_{c1}^2$. whereas the advection time across the shock precursor is $t_\mathrm{adv} \sim \kappa/v_1 c_{s1}$. The ratio of these two time scales is $t_\mathrm{grow}/t_\mathrm{adv}\sim \mathcal{M} \gg 1$, i.e. there is insufficient time for the instability to grow. 

We can understand the instability of the intermediate branch as follows, which is in line with suggestions that the divergent behavior of the intermediate branch is caused by a feedback loop between downstream CR pressure and acceleration efficiency \citep{drury81}. A clear criterion for a stable solution is $\partial P_{c2}/\partial P_{c1} > 0$, so that for instance the downstream CR pressure $P_{c2}$ decreases if the upstream value $P_{c1}$ decreases. Otherwise, the acceleration efficiency is divergent. In our variables, the stability criterion is: 

\begin{equation}
    \frac{\partial P_{c2}}{\partial Q} > 0 \Rightarrow \frac{\partial [\Delta P_{c}/(\rho_1 v_1^2)]}{\partial Q} > - \frac{P_{g1} + P_{c1}}{\rho_1 v_1^2} \approx - \frac{1}{\mathcal{M}^{2}} 
\end{equation}

where $1/{\mathcal{M}^{2}} \ll 1$. From Fig. \ref{fig:sol_struct}, we see that the strong negative slope of the intermediate branch implies that it is unstable, while both of the other branches are stable. Thus, a solution on the intermediate branch will evolve to one of the other branches. The instability of the middle branch in an 'S' shaped phase plane curve is generic to many problems, from thermal instability \citep{field69} to accretion disk instabilities \citep{smak84}. 

\begin{figure}
    \centering
    \includegraphics{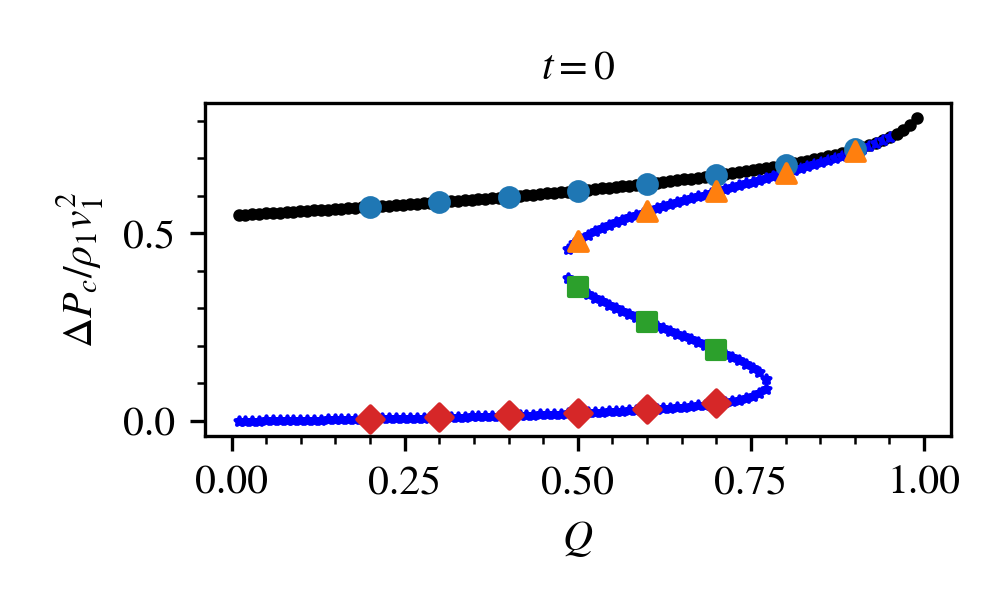} \\
    \includegraphics{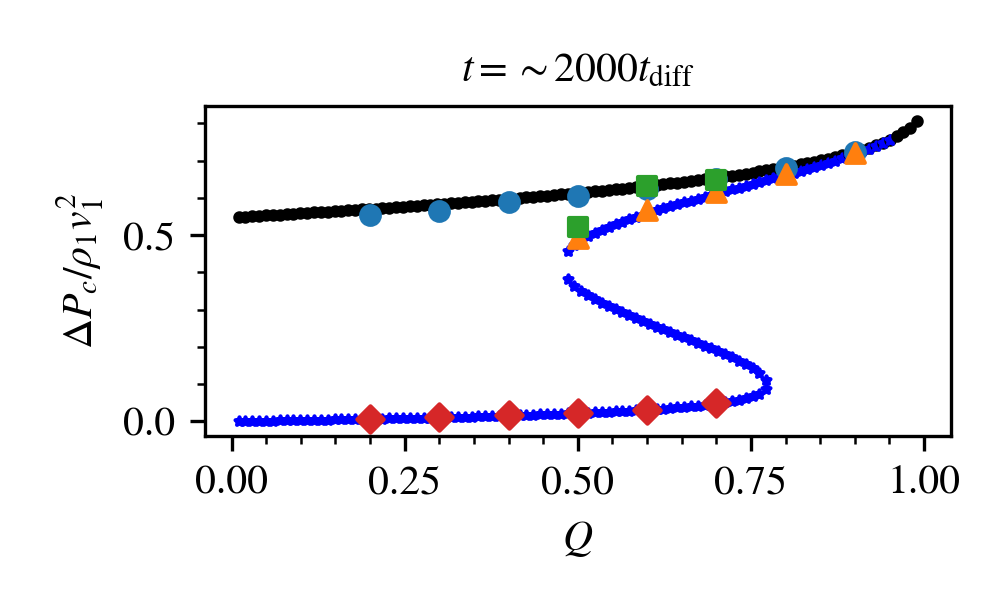}
    \caption{Acceleration efficiency against $Q$ for $\mathcal{M}=20, \beta=1$ at two time instances, at $t=0$ (Top) and $t\sim 2000 t_\text{diff}$ (Bottom). The markers denote the simulation data, the different marker shapes represents the solution branch the simulation was set up with (red diamond:  inefficient, green squares: intermediate, yellow triangle: efficient, blue circles: standard). The markers are threaded by black and blue lines, denoting the analytic acceleration efficiency (black corresponds to standard branch, blue corresponds to the new solution branches, i.e. efficient, intermediate, inefficient). The simulation data for the bottom plot are not all taken at the same time instance, they are taken at times that best represent the asymptotic behavior within simulation limits, at $t\sim 2000 t_\text{diff}$.}
    \label{fig:multiple_q}
\end{figure}

We proceed to test the analytic shock profiles for other values of $Q$. The result is shown in fig.\ref{fig:multiple_q}. The acceleration efficiencies of the standard, efficient and inefficient branches are stable and remain close to their initial setups while that of the intermediate branch is unstable, and asymptotes to the standard and efficient branch. These results show that our two-moment code handles this demanding test well, and in agreement with analytic expectations. 

\subsection{Setup 2: Free Flow} \label{subsec:free}

Next, we show how initial conditions influence which solution branch is realized. We simulate a fluid moving supersonically towards a reflecting boundary on the right at high speed, as in a converging flow. This causes the fluid to shock. The initial flow is either uniform or has background gradients. The left boundary is set to outflow if the flow is uniform initially, or with linear extrapolation in the ghost zones otherwise. The initial flow speeds are listed in table \ref{tab:free} and $P_c$ is set by $Q$. The CR flux $F_c$ is determined by eqn.\ref{eqn:cr_flux}. 

For uniform flow, $\rho$ and $P_g$ are set to $1000$ and $1$ respectively. In the setup with initial gradients, all quantities except for $P_c$ remain constant. $P_c$ was set to be linear: 

\begin{equation}
    P_c\qty(x) = \qty(P_{c1} - P_{c0})\qty(1 - \frac{x}{x_\mathrm{left}}) + P_{c0}, \label{eqn:pc_profile}
\end{equation}

where the subscripts $0,1$ denote quantities at the left and right boundaries respectively. Equation \ref{eqn:pc_profile} determines the spatially varying CR pressure fraction $Q\qty(x) = P_c\qty(x)/\qty(P_c\qty(x) + P_g)$. The initial profile can equivalently be parametrized by $Q_0$ and $Q_1$, the non-thermal fraction at the left and right boundaries. Since the shock propagagates from right to left, at first $Q=Q_{1}$, which then declines to $Q=Q_{0}$ as the shock moves leftward. This configuration is not in equilibrium. We therefore include constant artificial source terms to maintain background equilibrium, though we have explicitly checked that the branch selected is independent of the source terms. We set $\kappa = 0.1$ and use $N=16384$ grid cells unless otherwise specified.

\begin{figure*}
    \centering 
    \subfigure[$\mathcal{M}=14.8$, $Q=0.2$ and $\beta=1$]{\includegraphics{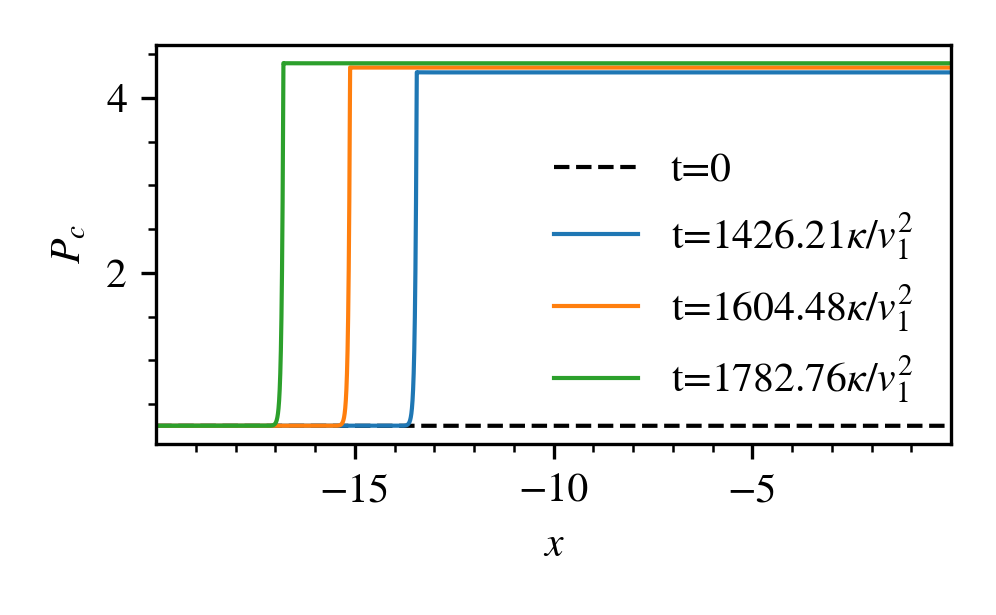}\label{fig:free.2}} 
    \addtocounter{subfigure}{2}
    \subfigure[Density]{\includegraphics{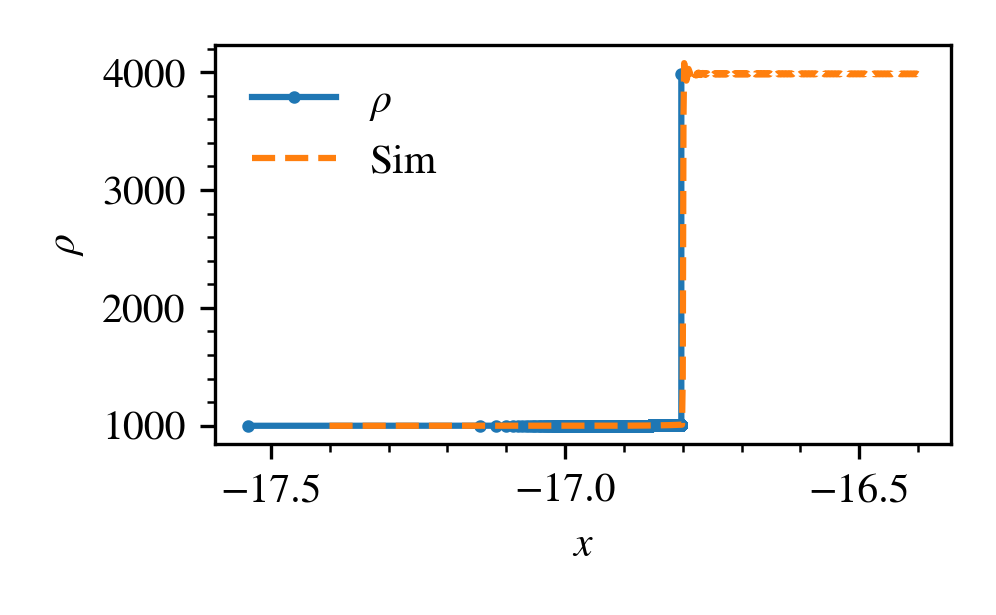}} \\
    \addtocounter{subfigure}{-3}
    \subfigure[$\mathcal{M}=26.6$, $Q=0.6$ and $\beta=1$]{\includegraphics{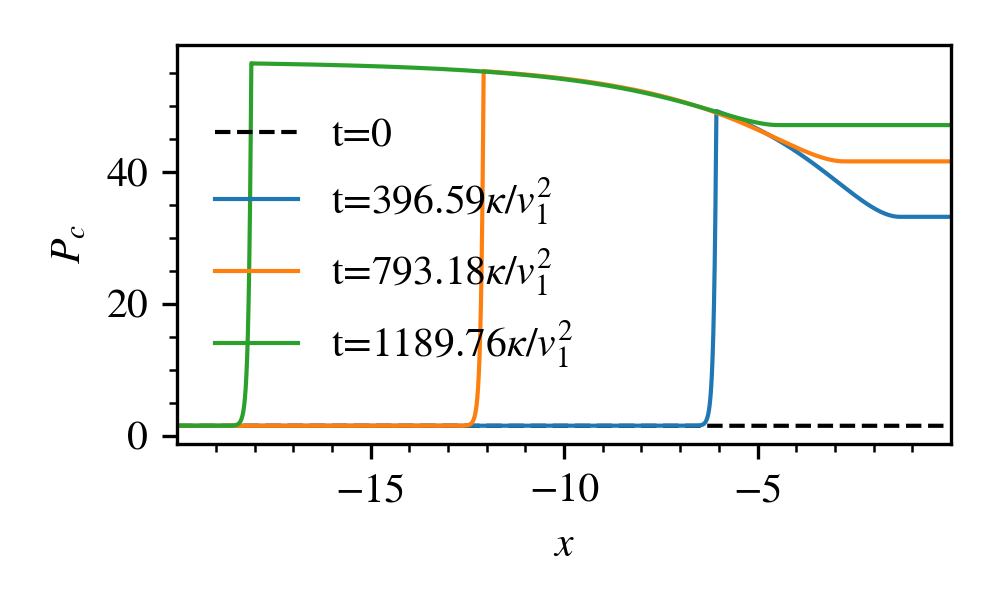}\label{fig:free.6}} 
    \addtocounter{subfigure}{2}
    \subfigure[Gas pressure]{\includegraphics{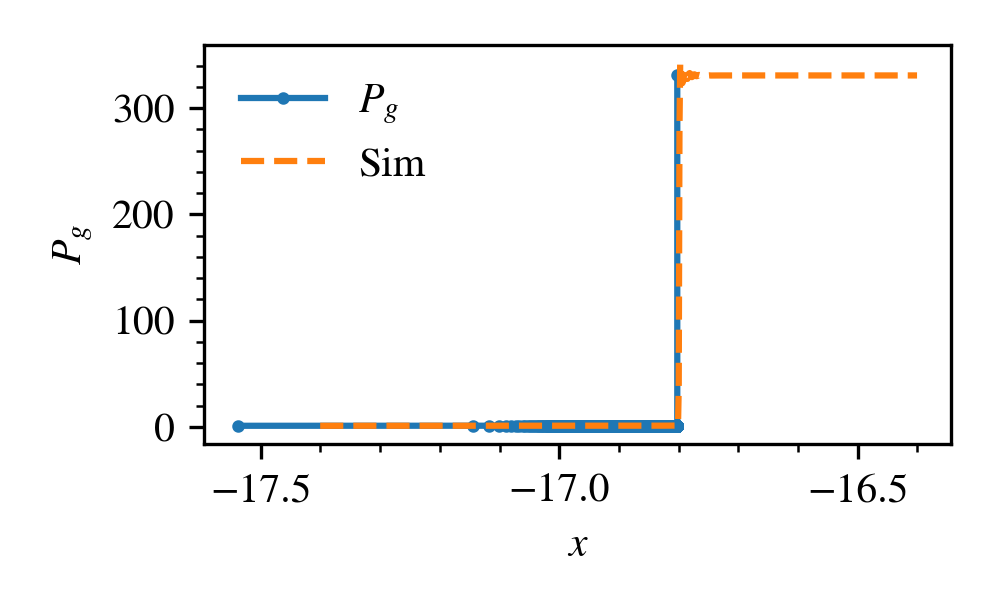}} \\
    \addtocounter{subfigure}{-3}
    \subfigure[$\mathcal{M}=22.4$, $Q=0.95$ and $\beta=1$]{\includegraphics{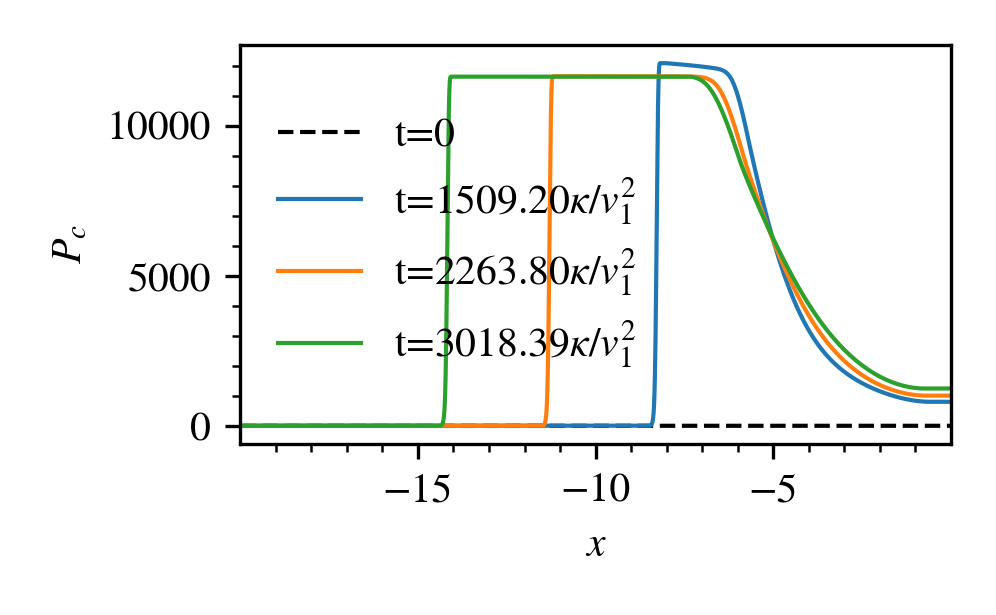}\label{fig:free.95}} 
    \addtocounter{subfigure}{2}
    \subfigure[Comparison of simulated shock profile against analytic for the $\mathcal{M}=14.8$, $Q=0.2$ and $\beta=1$ case.]{\includegraphics{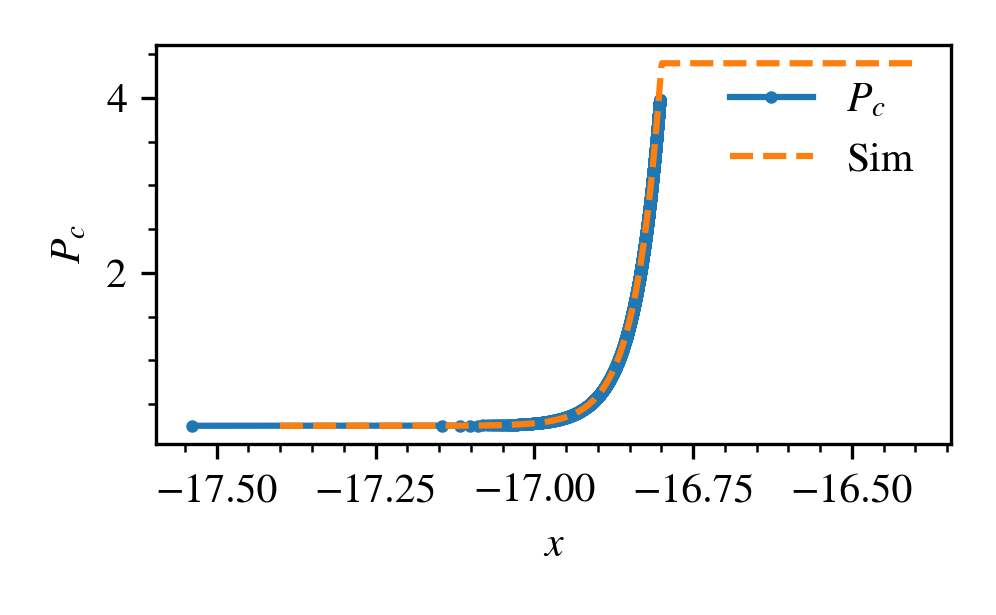}\label{fig:fmatch.2}}
    \caption{Simulation results for setup 2 without background gradients, as in Table \ref{tab:free}.}
    \label{fig:free}
\end{figure*}

\begin{table}
    \centering
    \begin{tabular}{c|c|c|c|c|c}
        \hline 
        \hline
        Case & v & $\qty(M,Q,\beta)$ & \# of & Selected & Acc. eff. \\
          & & & sol. branch & branch &  \\
        \hline
        Fig.\ref{fig:free.2} & 0.5 & $\qty(14.8,0.2,1)$ & 2 & Inefficient & 0.93\% \\
        Fig.\ref{fig:free.6} & 1.23 & $\qty(26.6,0.6,1)$ & 4 & Inefficient & 2.1\% \\
        Fig.\ref{fig:free.95} & 3.31 & $\qty(22.4,0.95,1)$ & 1 & Standard & 77.0\% \\
        \hline
    \end{tabular}
    \caption{Summary of test cases with uniform initial flow. The first column catalogs the corresponding figure. The second column lists the initial flow speed. The third column lists the upstream shock parameters. The forth column enumerates the number of analytical solution branches for each case. The fifth column records the branch selected by simulation. The last column measures the acceleration efficiency by $\qty(P_{c2} - P_{c1})/\rho_1 v_1^2$.}
    \label{tab:free}
\end{table}

For initially uniform flow, we show 3 different cases, which are tabulated in Table \ref{tab:free} and have profiles in Fig.\ref{fig:free.2}, \ref{fig:free.6} and \ref{fig:free.95}. The Mach number is measured in the shock frame with the shock velocity calculated by imposing continuity: $v_\mathrm{sh} = \qty[\rho v]/\qty[\rho]$. In each case, it took $\sim 1000 t_\mathrm{diff}$ for the shock to equilibrate. We comment further on these long equilibration times in \S\ref{subsec:further}. Equilibration generally takes longer for CR dominated flows. In fig.\ref{fig:free.95}, where there is a transition from low to high CR dominance, the equilibration time is extended by a factor of two.  We compare the simulated shock profile against the analytic prediction well after equilibration, and find good agreement. We show an example in fig. \ref{fig:fmatch.2}. This shows that bi-directional streaming is indeed necessary to understand the shock profiles. The slight discrepancy in the $P_c$ profile is due to fluctuations at the subshock associated with $\sign{\nabla P_{c}}$, as discussed in \S\ref{subsec:impose}. In Table \ref{tab:free}, it appears the inefficient branch is favored whenever it is a possible solution of the shock equations. We have found from all other shock simulations with uniform flow that the simulation indeed selects the inefficient branch whenever possible. It is unclear physically why this is the case, but could be related to the fact that the inefficient branch maximizes the wave entropy (the quantity on the LHS of equation \ref{eqn:adiabat}), and in particular has strongest subshock and hence the largest jump for the gas entropy. Consistent with the results of the previous section, the intermediate branch is never selected.

\begin{figure*}
    \centering
    \subfigure[The mach number $\mathcal{M}\approx 15$, $Q_0=0.2, Q_1=0.95$ simulation. The acceleration efficiency $\sim 57\%$, indicating the efficient/standard branch is selected.]{\includegraphics{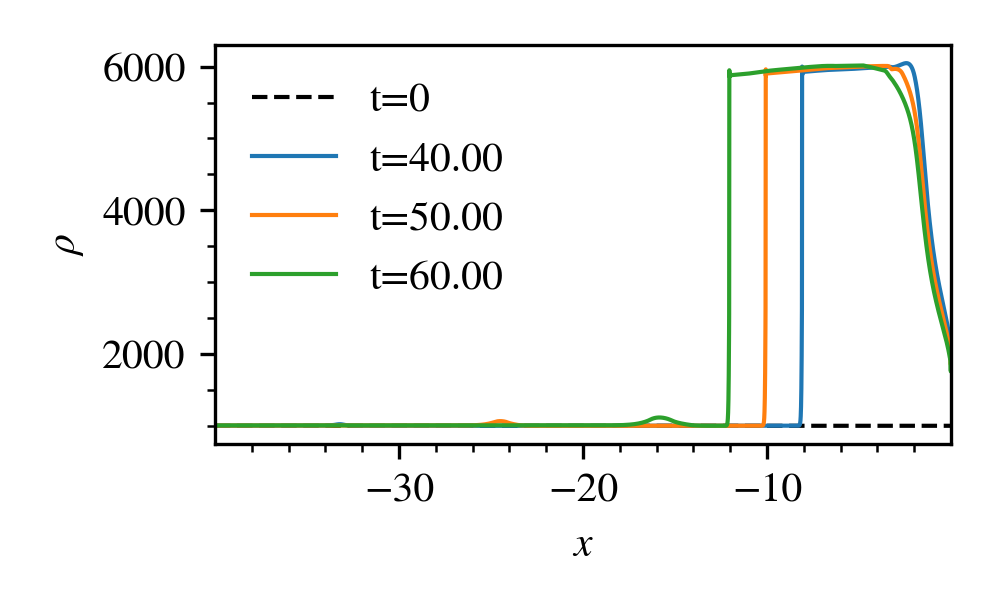}\includegraphics{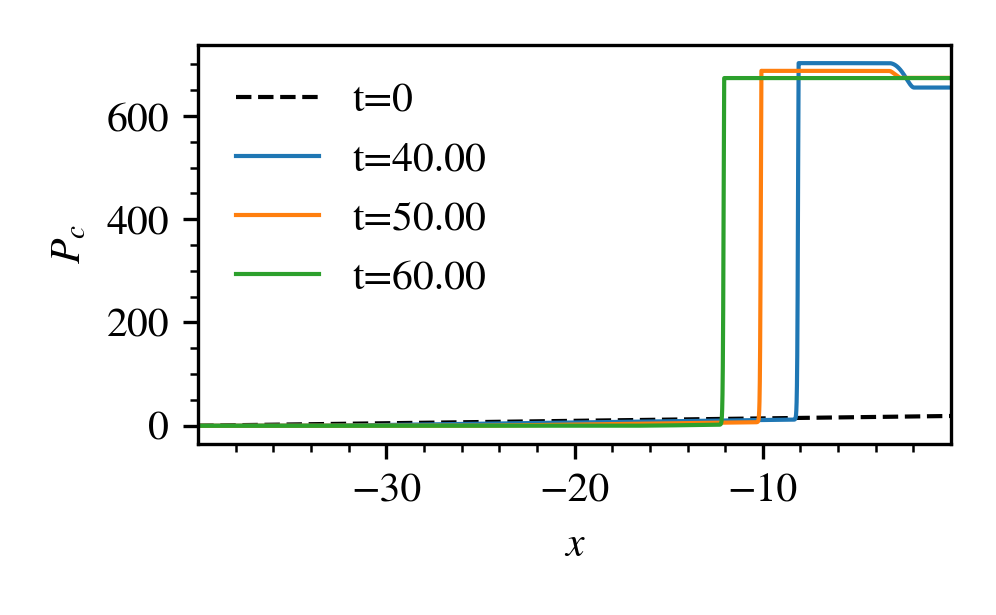}\label{fig:grad.95}} \\
    \subfigure[The mach number $\mathcal{M}\approx 20$, $Q_0=0.95, Q_1=0.75$ simulation. The acceleration efficiency transitions from $\sim 3\%$ to $\sim 77\%$, indicating the inefficient branch is selected at first, then the standard/efficient branch.]{\includegraphics{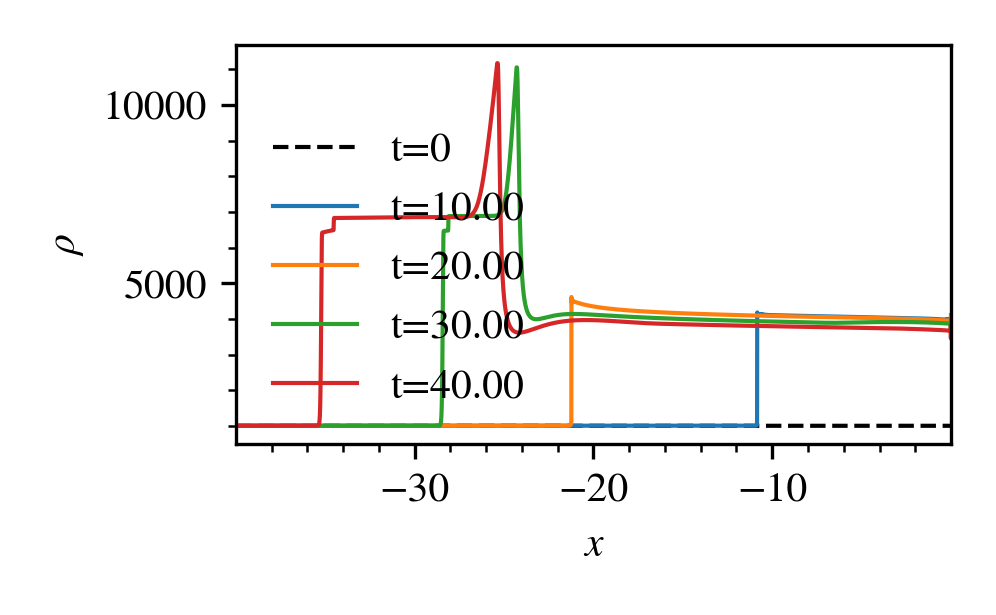}\includegraphics{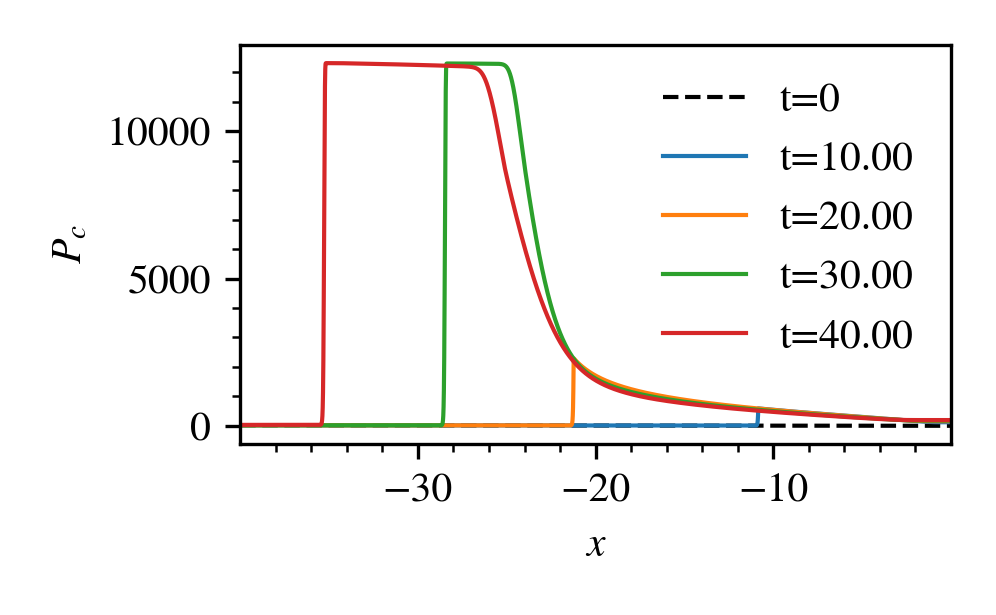}\label{fig:grad.75}}
    \caption{Simulation of setup 2 with background gradient. Time is given in code units because there isn't a well-defined $\kappa/v_1^2$ due to varying background.}
    \label{fig:my_label}
\end{figure*}

\begin{figure}
    \centering
    \includegraphics{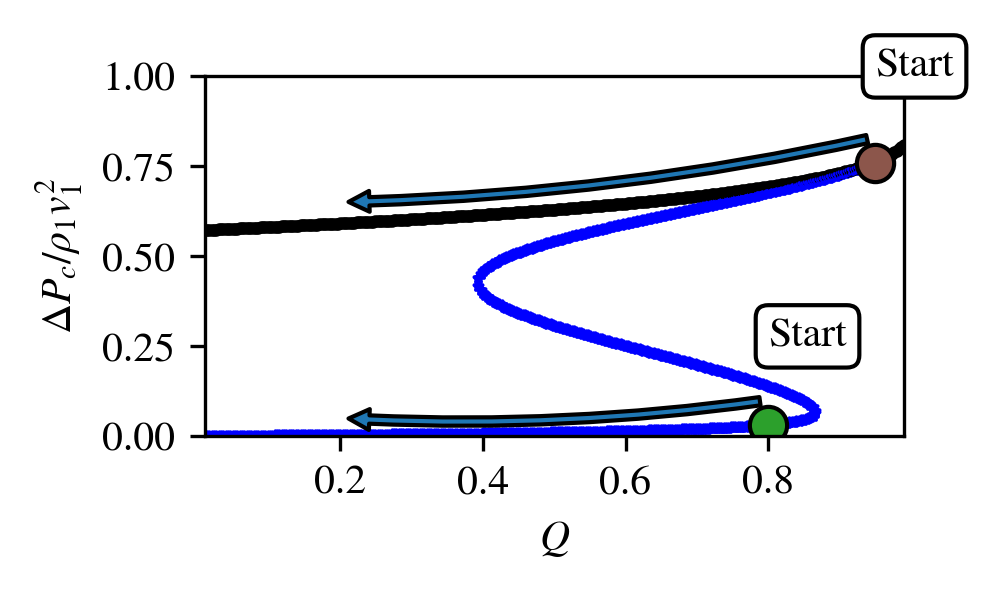} \\
    \includegraphics{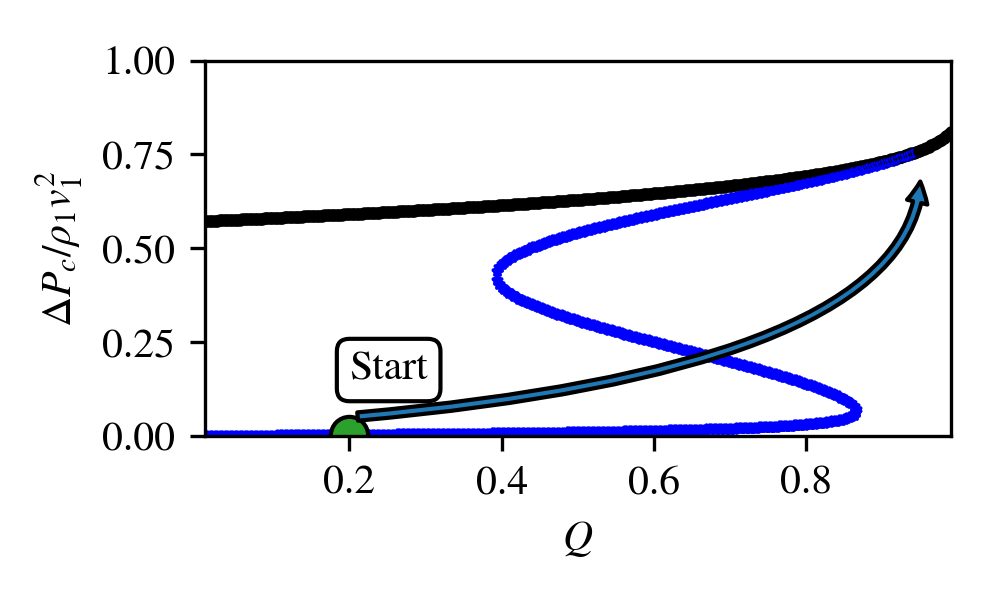}
    \caption{Acceleration efficiency against $Q$ plot showing how various solution branches are captured. \emph{Top}: A shock beginning at the efficient/standard branch (brown dot) would, under continuously varying background conditions, shift to another point on the same branch. The same holds for the green dot on the inefficient branch. \emph{Bottom}: A shock beginning at the inefficient branch (green dot) can transition to the efficient/standard branch if the background transitions into one for which only the efficient/standard branch is permissible. As before, the black solid line denotes the standard branch while the blue line denotes the new solution branches (efficient, intermediate and inefficient).}
    \label{fig:explain}
\end{figure}

Fig.\ref{fig:grad.95} shows the evolution of a shock in a background $P_c$ gradient, where $Q_0$, the initial non-thermal fraction at the left boundary, was $0.2$, while $Q_1$, that at the right boundary, was $0.95$. This was meant to simulate a shock propagating from a CR dominated region ($Q_1=0.95$), where only the efficient/standard branch is permissible, to a progressively gas dominated area, where the inefficient branch also exists. It is clear that the efficient/standard branch is picked throughout. As a comparison, a similar test where $Q_1=0.8$ was performed (not shown). The inefficient branch is selected throughout for this case. The reason for this discrepant solution pick is as follows: as shown in fig.\ref{fig:explain}, at $Q_1=0.95$, only the efficient/standard branch is possible, so the shock will pick this solution. Under continuously varying background conditions (in this case the gradually decreasing $P_c$), the shock will shift to a proximate point on the same branch. The shock remains on the same branch even if subsequent upstream conditions permit the inefficient branch. The same logic applies to the case $Q_1=0.8$. At $Q_1=0.8$, there are $4$ possible branches. As in our uniform background tests, the inefficient branch is picked. Subsequent evolution of the shock down the $P_c$ gradient follows the same branch. These two test cases have been repeated without the balancing source terms, causing the shock and the background to co-evolve with time. Nevertheless, the same result applies: the inefficient branch is selected for $Q_1 = 0.8$ while the efficient/standard branch is selected for $Q_1 = 0.95$. Branch selection is unaffected by source terms.

The reverse is also true. A shock beginning at the inefficient branch can, as in fig.\ref{fig:grad.75}, transition to the efficient/standard branch provided the upstream has shifted to conditions where only the efficient/standard branches are permissible. This could happen if the upstream is more CR dominated, or has higher plasma $\beta$. 

The findings of these simulations are summarized in Fig. \ref{fig:explain}. The branch selected by the shock is dependent on the local upstream conditions where the shock is formed. Where possible, the inefficient branch is picked. The shock will remain on the same branch unless the upstream transitions into conditions where only the efficient/standard branches are permissible. It will then switch to these branches and remain there. Thus, a shock passing through a CR dominated region (e.g., a cold cloud) will change its properties and continue to efficiently accelerate CRs, even after leaving the cloud. Two-fluid shock simulations appear to have hysteresis, likely because downstream conditions set boundary conditions for CR streaming which impact the shock itself. We will not consider the physics of this hysteresis, or the preference for the inefficient branch, further in this paper. The full realism of these properties is unclear, given the limitations of the standard two-fluid approach. For now,  it is important to be aware of them, given that two-fluid CR hydrodynamics is essentially the only approach used in galaxy formation simulations. 

\subsection{Setup 3: 1D Blast Wave} \label{subsec:blast}

Thus, far, we have focussed on the properties of steady-state shocks, and not examined properties of the time-dependent stage. We have already seen that the equilibration time of shocks can be long, $\sim 1000 t_{\rm diffuse}$. Thus, the acceleration efficiency of shocks will be time-dependent in a realistic setting. Here, as the simplest possible example, we consider a plane-parallel analog to a blast wave. 

In cosmological simulations, an SN event is typically prescribed to deposit mass, metals, momentum and energy to nearby cells of gas, generating an expanding shock wave. The energy deposited to CR (i.e. acceleration efficiency) is often taken to be $10\%$ of the total energy $\sim 10^{51}\ \mathrm{erg}$. If CR is treated as a fluid coupled to the thermal gas, additional CR will be generated at the expanding shock. As we have seen, this scan be handled self-consistently by a fluid code without a subgrid prescription, though whether the acceleration efficiency is correct as compared to PIC/hybrid simulations is another matter.

In our setup, a total energy of $E_\mathrm{ej} = 10^{51}\ \mathrm{erg}$ was deposited uniformly over a volume of radius $R = 10\ \mathrm{pc}$. $70\%$ of this was deposited into thermal energy, $10\%$ into CR energy and the remaining $20\%$ into kinetic energy. For a swept-up mass to be $50 M_\odot$, the average density was $\rho = 8.12\times 10^{-25}\ \mathrm{g}\,\mathrm{cm}^{-3}$. The average outflow velocity was therefore $v = 632\ \mathrm{km}\,\mathrm{s}^{-1}$, yielding a temperature of $T = 5.64\times 10^7\ \mathrm{K}$ from the ideal gas law for a gas of molecular weight $\mu = 1$. The surrounding ISM was assumed to have density $\rho_\mathrm{ISM} = 10^{-25}\ \mathrm{g}\,\mathrm{cm}^{-3}$ and $P_{g,\mathrm{ISM}} = P_{c,\mathrm{ISM}} = 10^3\, k_B\ \mathrm{K}\,\mathrm{cm}^{-3}$.  The Mach number of the expanding remnant is $\approx 40$. We consider both $\beta_\mathrm{ISM}=2,100$ cases. Following the analytic solution method described in \S\ref{sec:analytics}, there are 4 solution branches for the $\beta_\mathrm{ISM} = 2$ case, of which we expect the inefficient branch to be picked. For the $\beta_\mathrm{ISM} = 100$ case, only the efficient/standard branch is permissible. The whole domain spanned $-2000\ \mathrm{pc} < x < 2000\ \mathrm{pc}$, with outflow boundary conditions and $\kappa = 3\times 10^{25}\ \mathrm{cm}^2\,\mathrm{s}^{-1}$.  The acceleration efficiency is independent of the diffusion coefficient; the specific value chosen allowed the precursor to be resolved without an equilibration time which is too long (and requires a large box). A smaller diffusion coefficient is reasonable given the shorter mean free path of CRs at strong shocks, due to the amplification of magnetic perturbations. 
The domain was resolved with $N=65536$ cells (i.e. $0.06 \mathrm{pc}$ per grid cell). For simplicity and to avoid computational cost, the calculation is done in planar 1D geometry, and only meant to be illustrative.  

\begin{figure*}
    \centering
    \includegraphics[width=\textwidth]{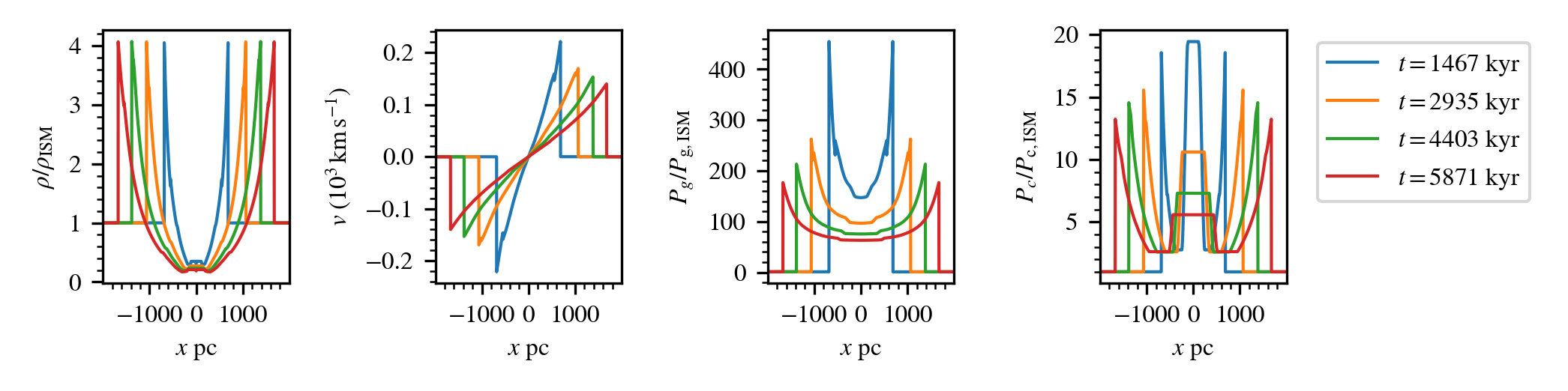} \\
    \includegraphics[width=\textwidth]{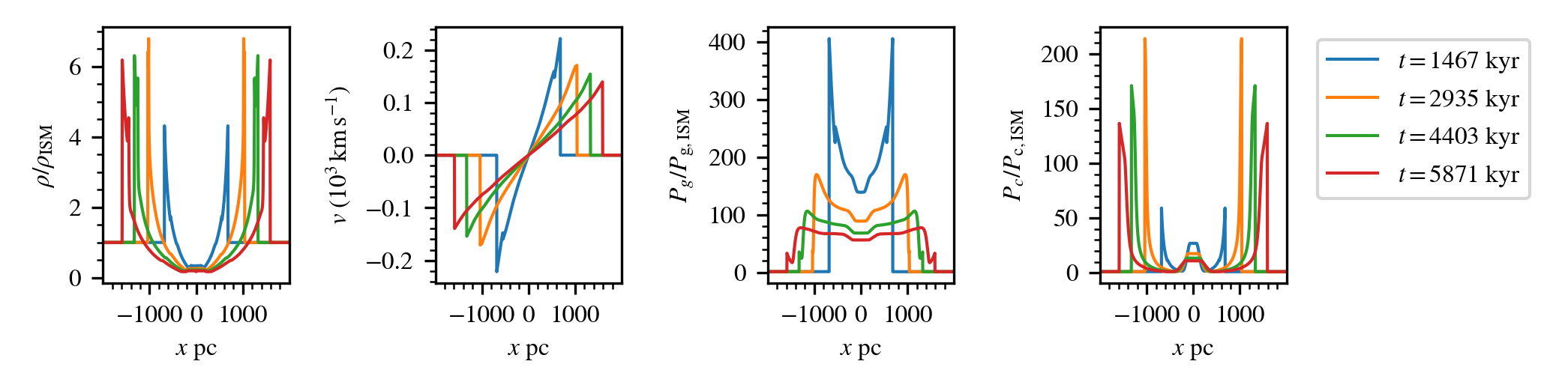}
    \caption{Evolution of 1D blast wave. \emph{Top row}: $\beta_\mathrm{ISM}=2$ \emph{Bottom row}: $\beta_\mathrm{ISM}=100$. The compression ratio and acceleration efficiency taken at $5871$ kyr are $\sim 4$ and $4.6\%$ for $\beta=2$ and $\sim 6$ and $67.3\%$ for $\beta=100$.}
    \label{fig:blast}
\end{figure*}

\begin{figure}
    \centering
    \includegraphics{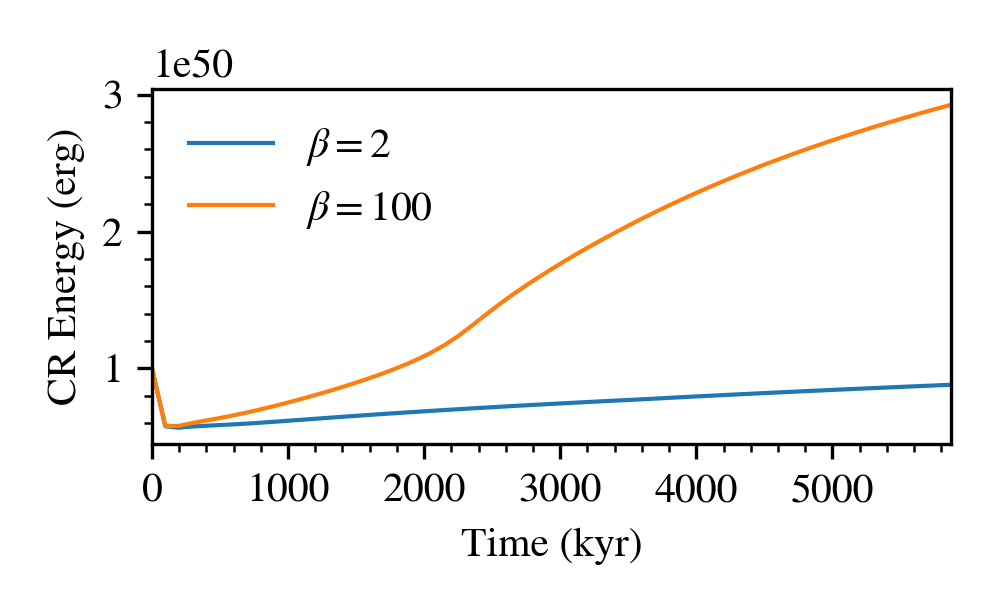}
    \caption{CR energy enclosed by the blast wave as a function of time for $\beta_\mathrm{ISM}=2$ (blue) and $\beta_\mathrm{ISM}=100$ (orange). The initial CR energy is $10^{50}$ erg, i.e. $10\%$ of the total energy ejected. After an initial transient phase, CR energy begins to rise due to particle acceleration.}
    \label{fig:creng}
\end{figure}

The top and bottom rows of fig.\ref{fig:blast} shows the time evolution of an expansion shock from a top-hat SNR setup for $\beta_\mathrm{ISM} = 2$ and $\beta_\mathrm{ISM} = 100$ respectively. After an initial transient of $\sim 1000$ kyr, the SNR settles onto a relatively stable structure. For $\beta_\mathrm{ISM} = 2$, the forward shock at $t=5871$ kyr has a compression ratio $\sim 4$ and an acceleration efficiency of $\sim 4.6\%$, indicating the inefficient branch is selected as expected. For $\beta_\mathrm{ISM} = 100$, the compression ratio rises to $\sim 6$ and the acceleration efficiency to $67.3\%$, indicating the efficient/standard branch is selected. As in fig.\ref{fig:free.95} in \S\ref{subsec:free}, the shock profile for the $\beta_\mathrm{ISM} = 100$ case underwent an extended transient of $\sim 1500$ kyr at low post-shock CR dominance before transitioning to the expected efficient/standard branch. Thus, the acceleration efficiency ramps up while the blast wave expands. Comparing the CR energy contained within the SNR of the two test cases (fig.\ref{fig:creng}), the high $\beta_\mathrm{ISM}$ case is clearly much more CR populated ($\sim 3.5$ times in this case). 

Our test case is clearly idealized and we expect the simulated profiles to change in realistic 3D spherical geometry, as well as the inclusion of additional physics such as radiative cooling and collisional losses. In particular, the forward shock should decelerate faster from stronger adiabatic cooling, reducing the acceleration efficiency and the net CR produced. Nevertheless, it shows how shocks can potentially add CRs over and above the initial values input by a subgrid recipe, as well as the influence of CR streaming losses (which differ in the low and high $\beta$ regimes) in reducing acceleration efficiency. 

\subsection{Further Considerations} \label{subsec:further}

Through most of this paper, we have considered time-steady, numerically resolved, parallel shocks only involving acceleration of pre-existing CRs (no injection from the thermal pool). Here, we briefly discuss the impact of relaxing these assumptions. 

\subsubsection{Long Equilibration Times} 

\begin{figure}
    \centering
    \includegraphics{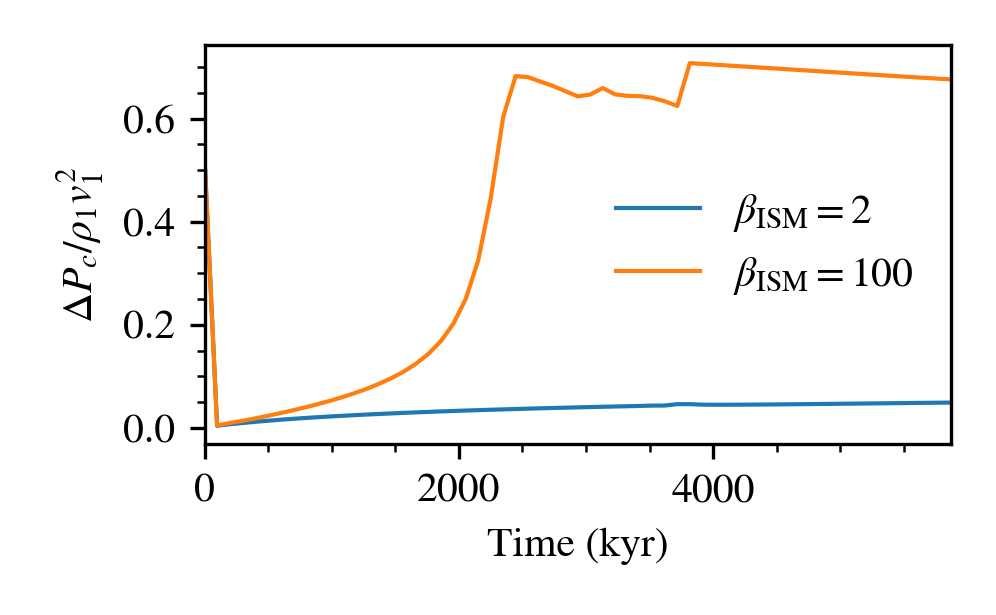} \\
    \includegraphics{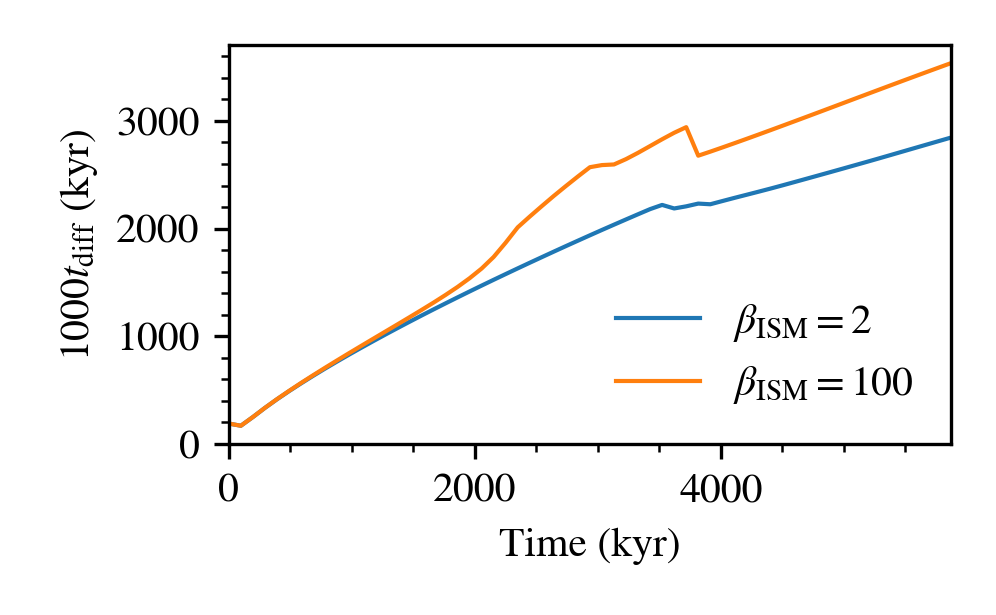}
    \caption{\emph{Top}: Acceleration efficiency as a function of time for the 1D blast wave example. \emph{Bottom}: Instantaneous diffusion time of the forward shock.}  
    \label{fig:time}
\end{figure}

We have already seen that shocks require $\sim 1000 t_\mathrm{diff}\sim 1000\kappa/v_1^2$ to equilibrate, where $t_\mathrm{diff}$ is the diffusion time and $v_1$ is the upstream velocity in the shock frame. The non-linear build up of the CR precursor, which significantly affects shock structure and CR acceleration, takes many CR diffusion cycles across the shock. The long equilibration time reflects the time required for the upstream flow to respond to the acceleration and diffusion of CR. This has been seen in previous work with diffusion only: for instance, \citet{jones90} found it took $\sim 200-1000 t_\mathrm{diff}$ for their solutions to equilibrate, which is very similar to our findings. Fig.\ref{fig:time} plots the acceleration efficiency and the instantaneous diffusion time of the forward shock in the setup described in \S\ref{subsec:blast}. Clearly, the equilibration time for the efficient/standard branch (the $\beta_\mathrm{ISM}=100$ case) is longer, $\sim 2500$ kyr. This timescale is indeed of order $\sim 1000 t_\mathrm{diff}$.

The equilibration time is longer for the efficient/standard branch because the post-shock CR pressure is higher, leading to a stronger precursor which takes a longer time to build up. By the same token, the more pre-existing CRs there are in the upstream, the more rapidly the precursor equilibrates. As mentioned in \S\ref{subsec:free} and \S\ref{subsec:blast}, when only the efficient/standard branch is permissible, there is usually an extended transient in which the shock transitions from low to high post-shock CR dominance (i.e. low to high CR acceleration efficiency), which coincides with the build-up of the precursor. This behavior was also seen by \citet{dorfi85, jones90} in simulations without CR streaming. \citet{jones90} derived an approximate analytic formula for the equilibration time and found that the number of diffusion time required is dependent on $\gamma_c$ as well. The equilibration time is the longest for $\gamma_c = 4/3$ and decreases for a stiffer CR equation of state, when the plasma is less compressible and the precursor plays a smaller role. For instance, in oblique shock simulations assuming $\gamma_c=4/3$, we find an equilibration time of $\sim 2500 t_\mathrm{diff}$ for $\theta=5\ \rm deg, \beta=100$ case, whereas  \citet{jun97} find $t_\mathrm{eq}\sim 36 t_\mathrm{diff}$ for $\gamma_c = 5/3$. Thus, in more realistic scenarios where $\gamma_c$ is self-consistently calculated (and varies continuously from $\gamma_c=5/3$ to  $\gamma_c=4/3$), equilibration times will be smaller. 

Nonetheless, the long equilibration times are important to keep in mind. Before reaching steady-state, shocks will have lower acceleration inefficiencies. One should be careful before grafting the result of steady state shock calculations in many astrophysical settings (for instance, when using a shock-finding algorithm to inject CRs by hand). For example, in SNR presented in \S\ref{subsec:blast}, the equilibration time is of order of 1 Myr, comparable to the expansion time, and the acceleration efficiency was clearly time-dependent. Other factors not present in our current simulations will affect whether the standard/high efficiency branch will appear in two-fluid galaxy formation simulations: by 1 Myr, radiative cooling will put the SNR in the snowplough phase, and various instabilities (e.g. Rayleigh-Taylor, CR acoustic instability, corrugational instability), if resolved, can disrupt the shock profile and truncate build-up of a precursor. 

\subsubsection{Numerical Resolution} 
\label{sec:resolution}

\begin{figure*}
    \centering
    \includegraphics[width=0.49\textwidth]{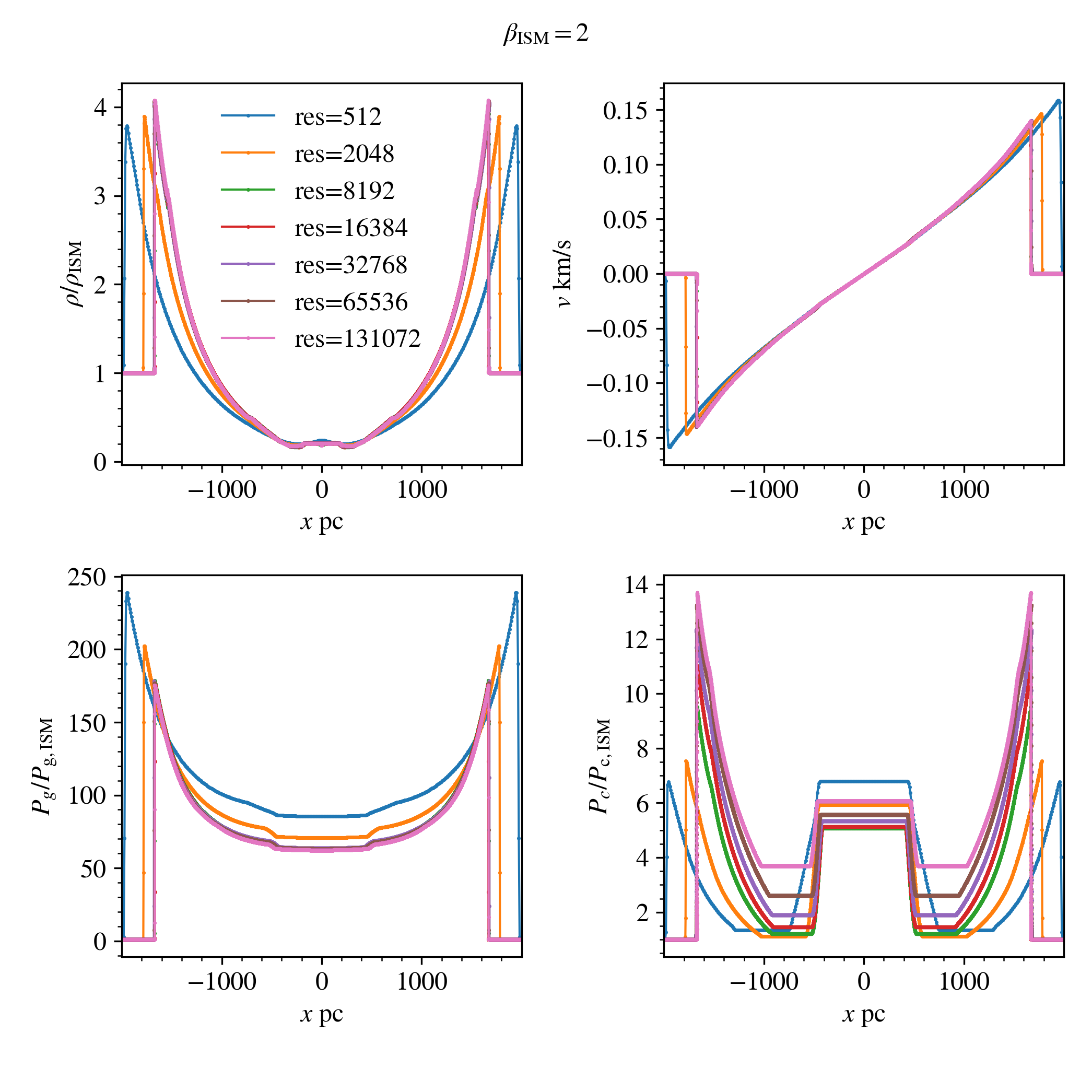}
    \includegraphics[width=0.49\textwidth]{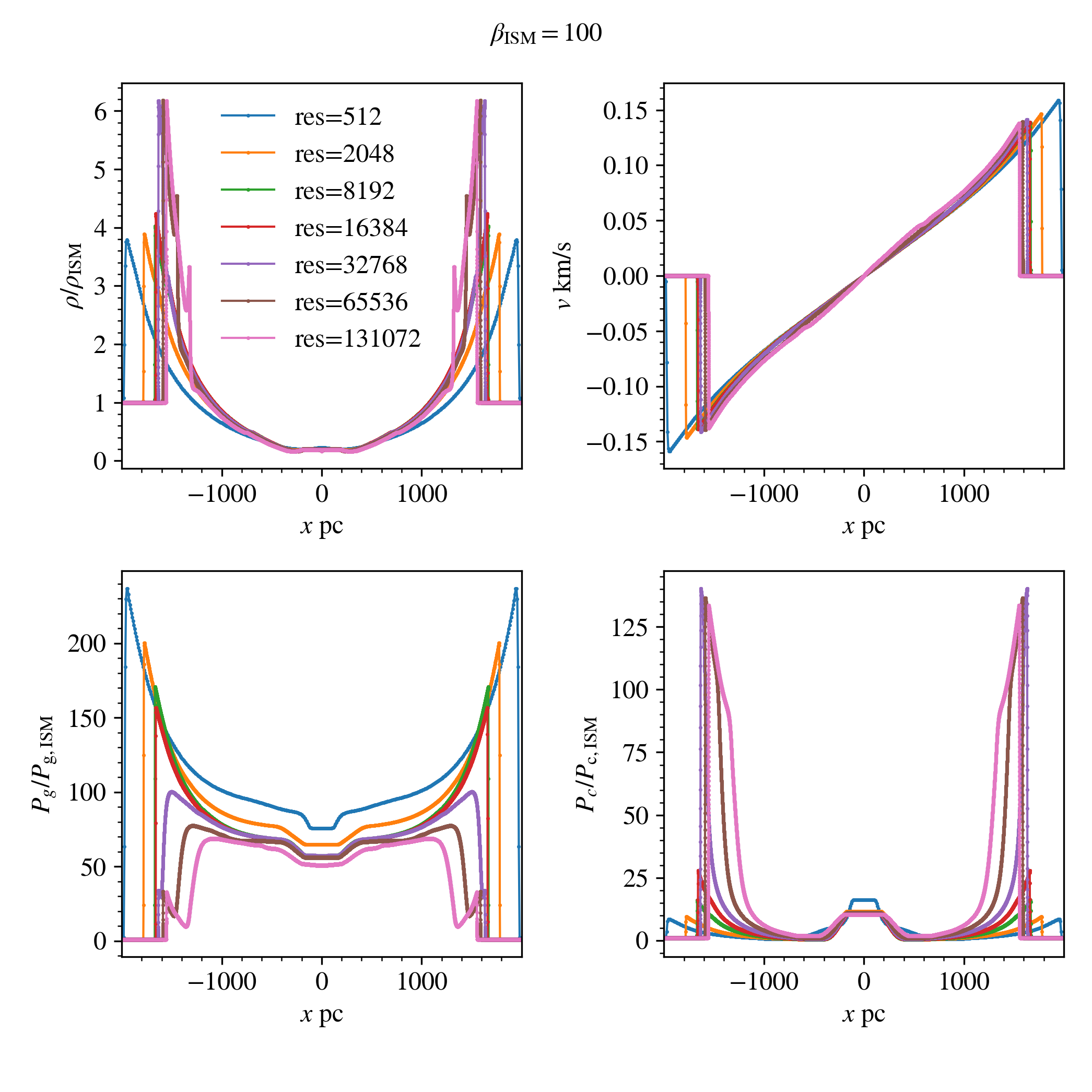}
    \caption{Plot of density, velocity, gas and CR pressure at $t=5871$ kyr at different resolutions for $\beta_\mathrm{ISM} = 2$ (left 4 panels) and $\beta_\mathrm{ISM} = 100$ (right 4). The legend indicates the number of grids used to resolve the simulation domain. Equivalently, the approximate number of grids the shock is resolved with is: 0.085 (res=512), 0.34 (res=2048), 1.37 (res=8192), 2.73 (res=16384), 5.46 (res=32768), 10.9 (res=65536), 21.9 (res=131072).}
    \label{fig:resolution}
\end{figure*}

\begin{figure*}
    \centering
    \subfigure{\includegraphics{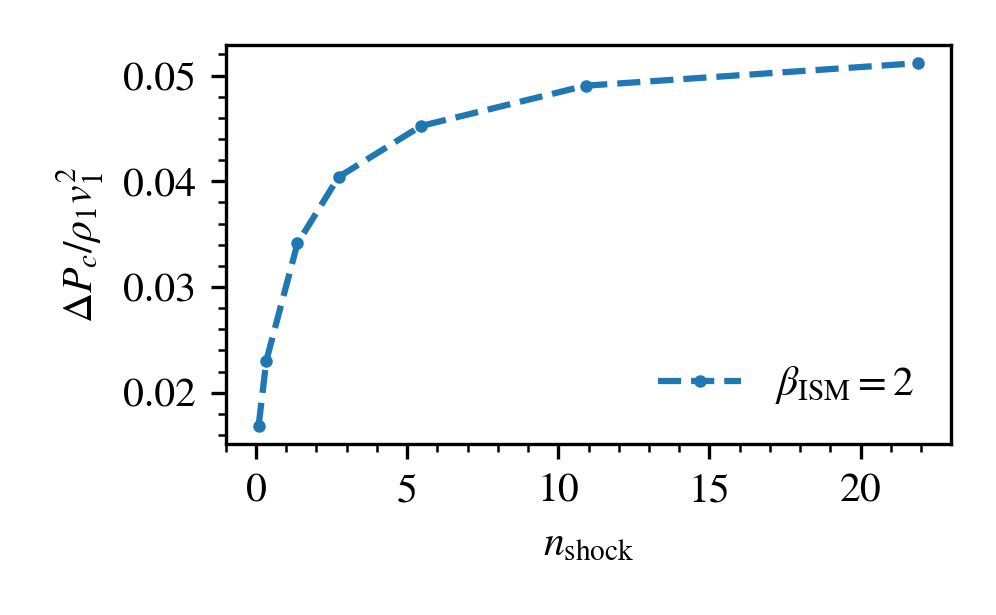}}
    \subfigure{\includegraphics{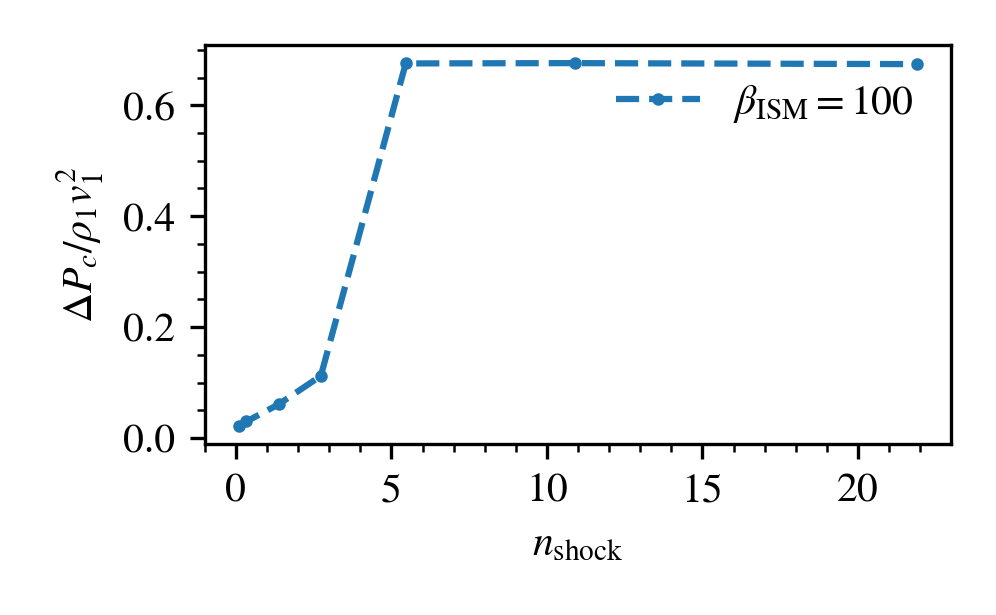}}
    \caption{Acceleration efficiency across the forward shock as a function of $n_\mathrm{shock}$ for $\beta_\mathrm{ISM} = 2$ (left) and $\beta_\mathrm{ISM} = 100$ (right) case.}
    \label{fig:acc}
\end{figure*}

It's clear that our high resolution simulations are converged, since they match analytic predictions. However, it is interesting to understand the minimal resolution needed to obtain accurate acceleration efficiencies. To study numerical convergence, we repeated the setup described in \ref{subsec:blast} at different resolutions, and compared solutions at $t=5871$ kyr. The number of grids used were: 512, 2048, 8192, 16384, 32768, 65536, 131072. Equivalently, taking the shock width at this time instance to be $\sim\kappa/v_1 = 3\times 10^{25}\ \mathrm{cm}^2\,\mathrm{s}^{-1}/150 \ \mathrm{km}\,\mathrm{s}^{-1} = 0.67\ \mathrm{pc}$ and a domain size of $4000\ \mathrm{pc}$ gives, in ascending order of resolution, the approximate number of grids $n_\mathrm{shock}$ the shock was resolved with: 0.085 (res=512), 0.34 (res=2048), 1.37 (res=8192), 2.73 (res=16384), 5.46 (res=32768), 10.9 (res=65536), 21.9 (res=131072). For $n_\mathrm{shock} < 1$, the shock is unresolved. One can see in fig.\ref{fig:resolution} that the solutions converges steadily for the $\beta_\mathrm{ISM} = 2$ case, whereas for the $\beta_\mathrm{ISM} = 100$ case, there is an abrupt transition from the inefficient branch to the efficient branch once $n_\mathrm{shock} > 5.5$. We quantified this by looking at the acceleration efficiency across the forward shock at different resolutions (fig.\ref{fig:acc}). The acceleration efficiency converges smoothly for the $\beta_\mathrm{ISM}=2$ case, while in the $\beta_\mathrm{ISM}=100$ case, there is slow change at low resolution followed by an abrupt rise at $n_\mathrm{shock}\sim 5$. 

Thus, the diffusion length must be resolved by $\sim 10$ grid cells for convergence in acceleration efficiency. At a lower Mach number, and if the upstream is highly CR dominated, the precursor is smaller and somewhat lower resolution may suffice. At insufficient resolution, the acceleration efficiency at shocks is underestimated. Except perhaps for very high resolution zoom simulations, most shocks in galaxy scale simulations will not resolve such length-scales and will thus have very low acceleration efficiencies. For this reason alone, it is likely safe to presume that the only source of CRs in such simulations are those injected by a sub-grid recipe.  

\subsubsection{Oblique Magnetic Fields} 
\label{sec:oblique}

\begin{figure*}
    \centering
    \includegraphics{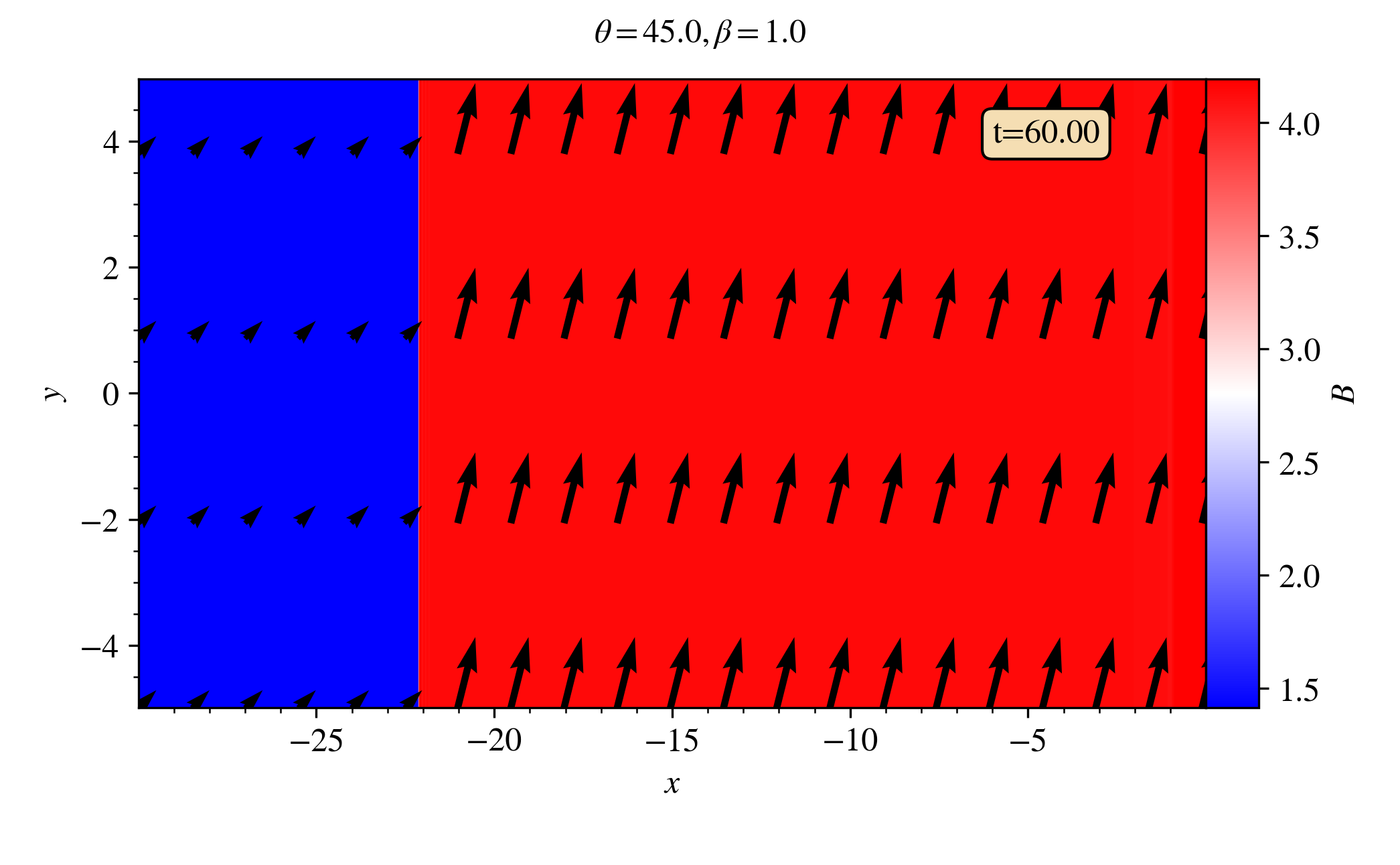} 
    \caption{The shock profile for $\theta=45\ \mathrm{deg},\beta=1$ at $t=60$ showing the  magnetic field. The arrows indicate the orientation of the field.}
    \label{fig:2d_beta1_b}
\end{figure*}

An oblique shock, where the magnetic field is no longer parallel to the shock normal, suppresses CR acceleration. This is because CR transport across the shock is suppressed.  In the post-shock fluid, compression preferentially amplifies the perpendicular B-field component, so the B-fields are aligned parallel to the shock front, suppressing  diffusion upstream. 

Here we describe four 2D test cases involving oblique magnetic fields. The setups were as follow: We initialized a uniform 2D flow of density $\rho=1000$, velocity $v=1.108$, gas pressure $P_g=1$ and CR pressure $P_c=1$ (i.e. $Q=0.5$) crashing towards the right boundary. The magnetic field was oriented at angle $\theta=5,45\ \rm deg$ to the shock normal for plasma beta $\beta=1, 100$. The reduced speed of light was set to $c=50$ and CR diffusivity to $\kappa_\parallel=0.1$ along the magnetic field and $\kappa_\perp=1.67\times 10^{-9}$ perpendicular to it. The domain spanned $-30 < x < 0$, $-5 < y < 5$ for the $\beta=1$ case and $-90 < x < 0$, $-5 < y < 5$ for the $\beta=100$ case. The whole domain was resolved with $2048\times 512$ grids for $\beta=1$ and $8192\times 512$ grids for $\beta=100$, corresponding to a precursor resolved by $n_\mathrm{shock}\approx 6,8$ grid cells respectively. Reflecting boundary was set at the right and outflow at the left. 

\begin{table}
    \centering
    \begin{tabular}{c|c|c|c|c|c}
        \hline 
        \hline 
        $\theta$ (deg) & $t\ \qty(t/t_\mathrm{diff})$ & $\qty(\mathcal{M}_f,Q,\beta)$ & $r$ & Acc. eff. & $\theta_\mathrm{out}$ (deg) \\
        \hline
        5 & 60 (1299) & $\qty(7.6,0.5,1)$ & 4.0 & 1.1\% & 19.5 \\
        45 & 60 (1304) & $\qty(7.6,0.5,1)$ & 4.0 & 0.9\% & 76 \\
        5 & 200 (3435) & $\qty(7.7,0.5,100)$ & 6.8 & 81.1\% & 30.8 \\
        45 & 200 (4344) & $\qty(8.65,0.5,100)$ & 4.0 & 1.2\% & 76 \\
        \hline
    \end{tabular}
    \caption{Summary of the oblique shock parameters. Column 1: Angle between upstream magnetic field and shock normal. Column 2: Time of measurement in code unit and in unit of the diffusion time in parenthesis. Column 3: Upstream Mach number (defined relative to the fast magnetosonic speed), non-thermal fraction and plasma beta. Column 4: Compression ratio. Column 5: Acceleration efficiency. Column 6: Angle between downstream magnetic field and shock normal.}
    \label{tab:oblique}
\end{table}

A summary of the oblique shock results is given in table \ref{tab:oblique}. The magnetic field for an example is also shown in fig.\ref{fig:2d_beta1_b}. Shock compression deflects the magnetic field away from the shock normal and increases its strength. Compressive amplification of magnetic field is stronger for higher obliquity as only the perpendicular component is boosted. The acceleration efficiencies of the $\beta=1$ case are very low ($\sim 1.1\%$ for $\theta=5\ \rm deg$ and $\sim 0.9\%$ for $\theta=45\ \rm deg$). This is inconsistent with the analytic prediction by \citet{webb86} (that included oblique magnetic fields and CR diffusion but no streaming), which predicted an efficiency of $\gtrsim 50\%$. The reduction in acceleration efficiency seen in our simulations is caused by bi-directional streaming, with the inefficient branch being picked. For simplicity we eschew repeating the analytic calculation in \S\ref{sec:analytics} including oblique magnetic fields. 

The difference is more marked at different obliquity for $\beta=100$. At $\theta=5\ \rm deg$, the efficient/standard branch is recovered, achieving an efficiency $\sim 81\%$. The acceleration efficiency decreases drastically at $\theta=45\ \rm deg$ to $\sim 1.2\%$. This suggests that the inefficient branch may be more extensive at high obliquity in parameter space ($\mathcal{M},Q,\beta$) than in the 1D case. Given that oblique shocks are the most common case, we expect the inefficient branch to appear more commonly in cosmological simulations than expected from 1D analytics and simulations. 

\begin{figure}
    \centering
    \includegraphics{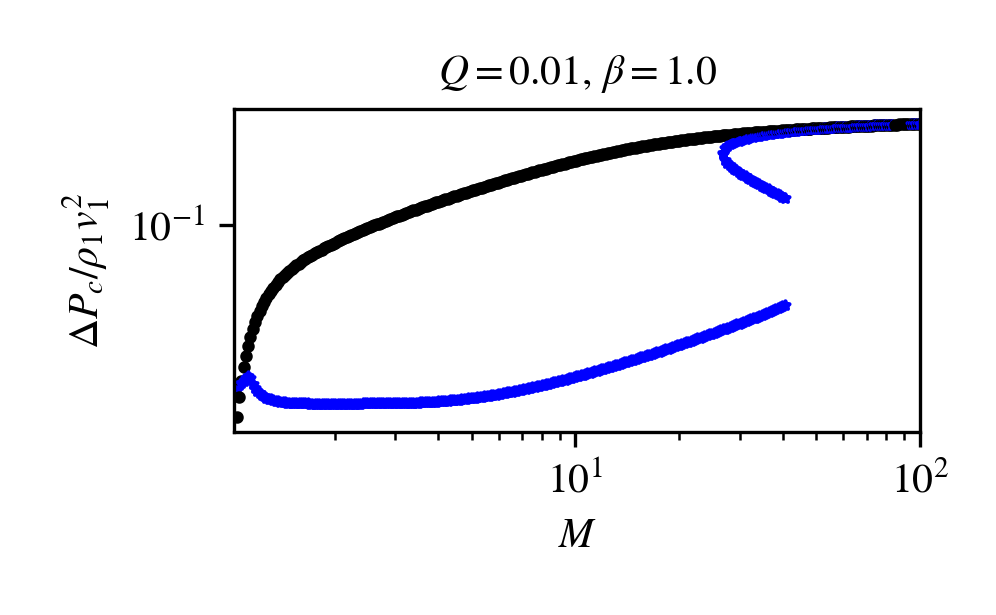}
    \caption{Acceleration efficiency against Mach number with injection. A fraction of $10^{-3}$ of the thermal particles is injected at the subshock. Injection is included only for the bi-directional solutions (blue curve). The uni-directional solution (without injection) is displayed here for comparison (black curve).}
    \label{fig:injection}
\end{figure}

\subsubsection{Injection of Thermal Particles} 
\label{sec:thermal} 

Thus far, we only consider acceleration of pre-existing CRs, which can easily take part in diffusive shock acceleration. However, suprathermal particles in the Maxwellian tail of the plasma can also be injected into the DSA process, and contribute to the CR population. This is particularly important when the pre-existing CR population is sparse (small $Q$). Here, we consider a simple prescription for injection which illustrates some potential effects. 

Following \citet{kang90}, injection can be accommodated in our solution method as follows. First, modify the subshock jump conditions to include injection:

\begin{gather}
    \qty[\rho v] = 0, \label{eqn:inj_mass} \\
    \qty[\rho v^2 + P_g] = 0, \label{eqn:inj_mom} \\
    \qty[\rho v \qty(\frac{1}{2}v^2 + \frac{\gamma_g}{\gamma_g - 1}\frac{P_g}{\rho})] = -I, \label{eqn:inj_eng} \\
    \qty[P_c] = 0, \label{eqn:inj_cr} \\
    \qty[F_c] = I, \label{eqn:inj_fc}
\end{gather}

where the injected energy flux $I$ is calculated according to

\begin{equation}
    I = \frac{1}{2}\epsilon\lambda^2 c_{s2}^2 J, \label{eqn:inj_I}
\end{equation}

for some prescribed injection efficiency $\epsilon$, postshock gas sound speed $\lambda c_{s2}$, and mass flux $J=\rho v$. Physically, this represents a small fraction $\epsilon$ of the incident thermal particles which have $\lambda$ times the postshock gas sound speed and are injected into the DSA process, contributing to the CR pressure. From the modified jump conditions, one can derive the relation:

\begin{equation}
    \qty(1 + \delta_\mathrm{inj}) J \langle v \rangle = \gamma_g \langle P_g \rangle, \label{eqn:inj_subsh}
\end{equation}
where

\begin{equation}
    \delta_\mathrm{inj} = \frac{\gamma_g - 1}{2}\epsilon \lambda^2\frac{c_{c2}^2}{\langle v \rangle\abs{\Delta v}}. \label{eqn:inj_delta}
\end{equation}

which is consistent with our previous equation \ref{eqn:jump} , if $\delta_\mathrm{inj}=0$. The symbols $\langle \rangle$ and $\Delta$ denote the arithmetic mean and the difference of the relevant quantity before and after the jump.

Graphically, equation \ref{eqn:inj_subsh} modifies the sonic boundary by a factor of $\qty(1 + \delta_\mathrm{inj})$. However, $\delta_\mathrm{inj}$ is not known a priori, so one must guess a value for $\delta_\mathrm{inj}$ first, then iteratively, using the updated pre and post subshock quantities, find an improved solution until the downstream state stays the same within some tolerance (taken to be $10^{-4}$ here). A good initial guess would be $\delta_0=\epsilon\lambda^2/2$.

We inject a fixed fraction $\epsilon \sim 10^{-3}$of the thermal particles into the CR population at the subshock, which is roughly the fraction of particles in a Maxwellian with $\lambda \sim 3$ times the sound speed. This fraction is consistent with the injection parameters in \citep{caprioli14a}. Fig.\ref{fig:injection} shows a case where most of the CRs come from injection ($Q=0.01$). Two points are worth noting: 1. The acceleration efficiency of the bi-directional solution increases with Mach number instead of the other way around. 2. At high Mach number the inefficient branch vanishes. The acceleration efficiency of the bi-directional solution ($\sim 5-10\%$) appears quite consistent with that found in hybrid simulations \citep{caprioli14a}. Our current simulation code not yet include injection; it is left for future work. However, it will improve the realism of two fluid shocks. We believe that most of the properties we have found (in particular, the fact that the inefficient bi-directional branch is favored) will continue to be found in simulations with injection. 

\section{Discussion and Conclusion} \label{sec:conclusion}

In this work, we studied steady-state CR modified shocks in the two fluid approximation, with the inclusion of both CR diffusion and streaming in the CR transport. This is a demanding test of new two-moment CR codes \citep{jiang18} which are the first to be able to handle such shocks with CR streaming; they have never been compared against analytic solutions. It also allows us to understand and quantify the effects of CR modified shocks in galaxy formation simulations. In a two-fluid code, shocks can accelerate CRs, over and above CRs injected via a sub-grid prescription; it is important to understand their contribution quantitatively. We only consider acceleration of pre-existing CRs, although one can modify the code to include thermal injection. Our findings are as follows: 

\begin{itemize} 
\item{{\it New analytic solutions: bi-directional streaming.} Previous analytic solutions \citep{volk84} assumed uni-direction streaming of CRs toward the upstream. In fact, over-compression at the subshock can lead to a transient spike (similar to the Zeldovich spike in radiative shocks) which seeds bi-directional streaming. The upstream and downstream CRs stream in opposite directions, away from the subshock. We obtain analytic profiles for this new solution. Streaming leads to lower acceleration efficiency with increasing magnetic field (due to increased gas heating and reduced compression). Furthermore, the new solution has a lower acceleration efficiency compared to the standard streaming, since downstream CRs propagate away rather than diffusing back to the shock.  The CR precursor is smaller and less compressible. At Mach number $\mathcal{M} \gtrsim 15$, the new solution bifurcates into inefficient, intermediate and efficient acceleration efficiency solution branches (fig. \ref{fig:mach} and \ref{fig:sol_struct}). The inefficient branch is a hydrodynamic shock only weakly modified by CRs, with acceleration efficiencies that typically do not exceed $10\%$. The efficient branch is CR dominated, with typical acceleration efficiency $\gtrsim 60\%$, similar to the standard branch. The intermediate branch lies somewhere in between. For weaker magnetic fields (higher $\beta$), the standard and new solutions merge closer together. At $\beta\gtrsim 100$ essentially only the efficient branch is left.}

\item{{\it Simulations match analytic solutions.} The simulations reproduces the standard analytic solution as well as all 3 branches of the new solution. The predicted acceleration efficiency also agrees extremely well with analytic predictions (fig.\ref{fig:fmatch.2}). It is excellent news  that the two-moment method can pass this demanding test, which should lay to rest concerns about solution degeneracy and numerical robustness at CR shocks \citep{kudoh16, gupta19}. As long as explicit diffusion is included (and for Fermi acceleration to operate, diffusion {\it must} be present), the analytic solution does not require ad hoc closure relations. As long as the diffusion length is resolved, numerical simulations closely match analytic solutions across a wide range of parameters.}   

\item{{\it Inefficient Branch Favored.} Which of the various solution branches is actually realized in nature? The intermediate branch is unstable (perturbations cause the acceleration efficiency to diverge to either the inefficient or efficient branch), so it is not realized. Of the remaining two possibilities, the branch selected is dependent upon the local upstream conditions where the shock is formed. In CR dominated shocks, only the efficient/standard branch is possible, since the compression ratio is high. However, if both branches are possible, the inefficient branch is selected, though transition to the efficient branch is possible if the upstream condition shifts to one for which only the efficient/standard branch is permissible. Once the shock selects the efficient/standard branch, it will remain there. See Fig. \ref{fig:explain}. The reason for this preference for the inefficient branch is unclear, though it is worth noting that it maximizes entropy generation at the shock (see discussion for diffusion only case in \citealt{becker01}).} 

\item{{\it Assumptions of time-steady, resolved and parallel shocks often not satisfied.} These calculations focus on well-resolved, steady-state, parallel shocks. These conditions are unlikely to be true in galaxy-scale simulations, and changes to these assumptions all point in the direction of reduced CR modification of the shock and lower acceleration efficiency: 
1. The equilibration time for a shock to reach its steady state structure is $t_{\rm equil} \sim 1000 t_\mathrm{diff}$ (where $t_\mathrm{diff}$ is the diffusion time); higher for CR dominated shocks with high acceleration efficiency, and somewhat smaller for shocks with lower acceleration efficiency. This is because the build-up of the CR precursor is a non-linear process which requires many diffusion times. This is often longer than shock crossing times (e.g. supernova remnants). Thus, shocks in realistic settings are not time-steady and cannot be compared directly to our results. 2. As shown in \S\ref{sec:resolution}, the precursor needs to be resolved by at least 10 grid cells for convergence. Lower resolution will lead to lower acceleration efficiency (fig.\ref{fig:acc}). 3. High obliquity magnetic fields will suppress formation of the CR precursor and hence acceleration efficiency, since the shock will more closely resemble a hydrodynamic shock with lower compression ratio (table \ref{tab:oblique}). 4. If thermal injection is taken into account,  the efficiency of the `inefficient' branch of the bi-directional solution ($\sim 5-10\%$) is in good agreement with hybrid/PIC simulations.}

\end{itemize}

In summary, in a two fluid code, the CR acceleration efficiency of shocks in a galaxy scale simulation is likely small ($\ll 10\%$) and thus the prevailing tendency to assume that they do not contribute significantly is likely reasonable. However, one must be careful to test this assumption, particularly in high resolution simulations, because the high efficiency branch converts such a large fraction ($\sim 60\%$) of the shock kinetic energy to CRs (e.g., see Fig \ref{fig:creng} where the CR energy rises far above the initial value), far above that obtained by kinetic simulations. In the end, we find that in most settings a two fluid code `does no harm' at shocks and gives roughly physical reasonable solutions, despite the significant shortcomings of the fluid approach in handling a fundamentally kinetic problem, as discussed in the Introduction. The fluid approach can probably be modified (e.g., introducing thermal injection, as in \S\ref{sec:thermal}, and potentially introduce a time-dependent $\kappa$ and $\gamma_c$ as the shock evolves) which further improves agreement with kinetic results. In the end, however, the most pragmatic approach for galaxy-scale simulations is to simply leave the code as-is, effectively ignoring CR injection at shocks. If the CR acceleration at shocks is a critical application, then one can simply apply a shock finding algorithm and inject CRs by hand (e.g., \citealt{pinzke13, pfrommer17}), but carefully taking the time-dependence of acceleration into account.  

\section*{Acknowledgements} 

We thank Omer Blaes and Eliot Quataert for helpful conversations. We acknowledge NSF grant AST-1911198 and XSEDE grant TG-AST180036 for support.  
The Center for Computational Astrophysics at the Flatiron Institute 
is supported by the Simons Foundation. 

\section*{Data Availability} 

The data underlying this article will be shared on reasonable request to the corresponding author. 



\bibliographystyle{mnras}
\bibliography{main} 






\bsp	
\label{lastpage}
\end{CJK*}
\end{document}